\documentclass{article}

\usepackage{amssymb,amsmath,amsthm}
\usepackage{color}
\usepackage[mathcal]{euscript}
\usepackage{microtype}
\usepackage{array}
\usepackage{graphicx}
\usepackage[margin=1in]{geometry}
\usepackage{hyperref}
\usepackage[hang,flushmargin,bottom]{footmisc}
\usepackage[arxiv]{optional}

\graphicspath{{./fig/}}

\makeatletter
\renewcommand\subparagraph{%
 \@startsection {subparagraph}{5}{\z@ }{3.25ex \@plus 1ex
 \@minus .2ex}{-1em}{\normalfont \normalsize \bfseries }}%
\makeatother

\usepackage{amsthm}
\usepackage{mathtools}
\usepackage[dvipsnames,usenames]{xcolor}
\usepackage{hyperref}
\hypersetup{colorlinks=true, urlcolor=Blue, citecolor=Green, linkcolor=Black, breaklinks, unicode}

\usepackage{float}
\usepackage{graphicx}
\graphicspath{{figures/}} 
\DeclareGraphicsExtensions{.pdf,.png,.jpg}

\usepackage{mathdesign}

\usepackage[nocompress]{cite}
\usepackage{enumerate}
\usepackage{paralist}
\usepackage{bm}

\usepackage[utf8]{inputenc} 
\usepackage[T1]{fontenc}    
\usepackage{hyperref}       
\usepackage{url}            
\usepackage{booktabs}       
\usepackage{amsfonts}       
\usepackage{nicefrac}       
\usepackage{microtype}      
\usepackage{xcolor}         

\usepackage{amssymb, amsmath, cleveref, amsthm, mathtools, graphicx, subcaption} 

\newtheorem{lemma}{Lemma}
\newtheorem{theorem}[lemma]{Theorem}
\newtheorem{definition}{Definition}

\newcommand{\recom}{\textsf{ReCom}}
\newcommand{\audit}{\textsc{LF Auditing}}
\newcommand{\generation}{\textsc{LFP Generation}}
\newcommand{\ens}{\mathcal{E}}
\newcommand{\ckspb}{\textsc{C$k$S-PB}}

\newcommand{\unf}{\text{unf}}

\def\eps{\varepsilon}

\def\abs#1{\left| #1 \right|}
\def\Paren#1{\left( #1 \right)}

\def\floor#1{\left\lfloor #1 \right\rfloor}

\newcommand*\samethanks[1][\value{footnote}]{\footnotemark[#1]}

\title{All Politics is Local: Redistricting via Local Fairness\thanks{This work is supported by NSF grants CCF-2113798, IIS-1814493, CCF-2007556, and CCF-2223870. }}

\author{
    Shao-Heng Ko\thanks{Equal contributions.} \qquad
    Erin Taylor\samethanks[2] \qquad
    Pankaj K. Agarwal \qquad
    Kamesh Munagala \\
  \small{Department of Computer Science}\\
  \small{Duke University}\\
  \small{Durham, North Carolina 27708 USA}\\ 
  \small{Email: \texttt{\{sk684,ect15,pankaj,kamesh\}@cs.duke.edu}}
}

\begin{document}

\maketitle

\begin{abstract}
In this paper, we propose to use the concept of {\em local fairness} for auditing and ranking redistricting plans. Given a redistricting plan, a \emph{deviating group} is a population-balanced contiguous region in which a majority of individuals are of the same interest and in the minority of their respective districts;
such a set of individuals have a justified complaint with how the redistricting plan was drawn. A redistricting plan with no deviating groups is called \emph{locally fair}. We show that the problem of auditing a given plan for local fairness is \textsf{NP}-complete. We present an MCMC approach for auditing as well as ranking redistricting plans. We also present a dynamic programming based algorithm for the auditing problem that we use to demonstrate the efficacy of our MCMC approach. Using these tools, we test local fairness on real-world election data, showing that it is indeed possible to find plans that are almost or exactly locally fair. Further, we show that such plans can be generated while sacrificing very little in terms of compactness and existing fairness measures such as competitiveness of the districts or seat shares of the plans. 
\end{abstract}

\section{Introduction}
\label{s:intro}

Redistricting in the United States is the process of partitioning a state into districts, each of which elects one representative to the Congress, for the most part, via simple majority voting.
As of April 2022, one year after the US Census Bureau released the results of the 2020 decennial census, 41 out of the 50 states have finished redrawing the congressional redistricting plans for the next decade~\cite{redistricting101}. This process has triggered numerous debates and litigation along the way. Much of this debate centers on whether the plans are \emph{gerrymandered} so that one of the two parties gets more representatives. Given its high-stakes impact and mathematical richness, there has been persistent interest in tackling redistricting as an algorithmic question since the early $1960$s~\cite{becker2020redistricting}. 

There is ongoing debate around what a ``desirable'' redistricting plan should be. It is commonly agreed that ``desirable'' plans should, at minimum, produce population balanced, contiguous, and compact districts~\cite{reynolds_sims}. Beyond this basic agreement, there is still debate on richer notions of desirability, particularly notions related to the ``fairness'' of a plan. This has motivated a long line of recent work~\cite{deford2019recombination, deford2020computational, herschlag2017evaluating} as well as software tools~\cite{princeton, redistricting101} on \emph{auditing} a given redistricting plan against various fairness concepts. Some of these concepts have since been adopted in Wisconsin's and Michigan's redistricting efforts~\cite{recomusage}. It should be noted that under most notions of desirability proposed in literature, the problem of redistricting is computationally hard~\cite{kueng19fair}, leading to the study of heuristic approaches, which we outline later.

\noindent {\bf Global versus Local Fairness.} Zooming into fairness criteria, most extant notions of fairness focus on the \emph{global outcomes} of the redistricting plans, e.g., whether the \emph{seat shares} proportionally represent the demographics~\cite{warrington2018quantifying}, or how competitive the districts drawn are~\cite{deford2020computational}. However, it is argued in~\cite{agarwal2021locally} that global metrics do not always distinguish between \emph{natural} gerrymandering -- when the distribution of voters unavoidably prohibits certain globally fair outcomes -- and \emph{artificial} gerrymandering -- when the plans are manipulated to favor a demographic group. This issue is typically addressed via {\em statistical tests}~\cite{deford2019recombination}: a probabilistic method is used to generate an ensemble of population balanced, contiguous, and compact plans, and the global fairness score in question is computed for each of these plans, yielding a histogram of scores. The plan in question is deemed ``fair'' if its global fairness score is not an outlier in this histogram. 
 
Global fairness, such as proportional seat shares, despite being desirable and statistically testable, may not represent the \textit{local} concerns of voters. For instance, imagine the blue party cares about rising sea levels and climate change, while the red party does not. In North Carolina, if at least one seat on the eastern coast has blue majority, that representative may advocate to mitigate the impacts of climate change to the coastal residents on the state or federal level. On the other hand, a better seat share may lead to a plan in which all districts near the coast have red majority, while the districts in the western mountains have blue majority. However, the latter set of representatives may not advocate for issues impacting the coastal residents, since it is not of local concern to the geographic area. This motivates the need for local fairness as a separate fairness measure, capturing at some level the saying ``all politics is local''.

Borrowing the notion of core from cooperative game theory, the work of~\cite{agarwal2021locally} defines local fairness notion as follows: given a redistricting plan, a voter is unsatisfied if the majority demographic in her district does not match her own demographic. A redistricting plan is locally fair if no group of unsatisfied voters could \emph{deviate} and draw a different district such that this group of unsatisfied voters has a majority in the new district. 

As in the scenario above, such a local notion of fairness has the advantage of capturing \emph{justified complaints} of groups of voters, as has happened in earlier court judgements~\cite{cooper_harris}. It also provides a way of auditing enacted plans without resorting to statistical tests, making it more human interpretable and {\em explainable}. 

\noindent {\bf Research Questions.} The notion of local fairness is appealing; however, the analysis and results in~\cite{agarwal2021locally} are theoretical and apply only to a simplified one-dimensional model. In this paper, we develop algorithms to audit plans for local fairness, and systematically study this concept on real-world electoral data. In particular, we study the following questions:

\begin{itemize}
    \item Given a redistricting plan, can we efficiently test (or audit) whether the plan is locally fair?
   \item Are locally fair plans achievable in real redistricting tasks? If not, can we quantify how far a given plan is from being locally fair? 
  \item Is local fairness empirically compatible with other existing global fairness concepts? 
  \end{itemize}

\subsection{Related Work}
\label{ss:related}
 \noindent {\bf Redistricting as Optimization.}
We first focus on the task of drawing plans, or computational redistricting. The idea of using computational tools in redistricting dates back to the 1960s~\cite{vickrey1961prevention, hess1965nonpartisan}. Since then, an extensive line of work (see~\cite{becker2020redistricting} for a comprehensive survey) cast the redistricting task as an optimization problem, in which the input contains only spatial location of individuals, but not their political affiliations. The objective and constraints capture the population balance, contiguity, and compactness criteria of the districts. This problem is computationally intractable in the worst case~\cite{cohen-addad2021computational}, and multiple algorithmic approaches have been proposed, including Voronoi diagrams~\cite{levin2019automated, gawrychowski2021voronoi}, local search~\cite{king2015efficient}, simulated-annealing and hill climbing~\cite{altman2011bard}, and spatial evolutionary algorithms~\cite{liu2016pear}.  On the flip side, it is argued in~\cite{chen2013unintentional, wheeler2020impact} that such ``neutral'' districting plans -- as outputs of algorithms without political inputs -- may contain unintentional biases, as well as unexpected outcomes such as ``natural gerrymandering''~\cite{borodin2018big, garg2021combatting}, i.e., the geographic distributions of voters naturally lead to disproportionate seat shares. Therefore, fairness objectives such as partisan representativeness are typically incorporated into the redistricting problem as objectives; however, these additional requirements further add to the computational difficulty of the problem~\cite{kueng19fair}.

 \noindent {\bf Ensemble Approaches to Redistricting.}
Instead of optimizing and finding a single best redistricting plan, another line of work focuses on generating a large \emph{ensemble} of districting plans, with the hope of some of these plans being fair. These methods include Flood Fill~\cite{cirincione2000assessing, magleby2018new}, Column Generation~\cite{gurnee2021fairmandering}, and the widely adopted Markov Chain Monte Carlo (MCMC) approach~\cite{tam2016toward, fifield2020automated, lin2022auditing}. The latter approach \emph{samples} from the space of feasible plans with a bias towards ``desirable'' or fairness properties.  For instance, it is shown in~\cite{procaccia2022compact} that the widely used \recom\ MCMC method~\cite{deford2019recombination} provably biases towards compact plans.  We provide a more in-depth description of \recom\ in Section~\ref{s:algorithm}. The work of~\cite{esmaeili2022centralized} proposes a method for choosing one representative plan from such an ensemble based on defining distances between plans.

 \noindent {\bf Auditing and Combating Gerrymandering.}
A somewhat different question from constructing a desirable plan is the question of {\em auditing} a given plan for desirability and fairness. As mentioned before, ensemble based approaches provide a natural, statistical way of auditing~\cite{herschlag2017evaluating, herschlag2020quantifying}: The properties of the enacted plan is compared against the histogram of the corresponding property on the ensemble; if the plan is a statistical outlier, then it is considered more ``gerrymandered'' and hence less desirable. The recent work of~\cite{lin2022auditing} instead uses plans in the ensemble as comparators to identify manipulation in redistricting plans. On the non-statistical side, numerous approaches to auditing have also been proposed via appropriate desirability scores. These are either scores based on compactness of the plan (such as the Reock~\cite{reock1961note} and Polsby-Popper~\cite{polsby1991third} scores), or scores based on partisan outcomes generated by the plan (such as the efficiency gap~\cite{stephanopoulos2015partisan}, mean-median gap~\cite{wang2016three}, partisan symmetry~\cite{warrington2018quantifying}, and the GEO metric~\cite{campisi2022geography}), or scores based on competitiveness of the plan~\cite{deford2020computational}. Many of these measures are used in publicly available tools~\cite{princeton, GEO}. Finally, there is a recent line of work that attempts to eliminate gerrymandering by completely revamping the winner-takes-all, single-member district mechanism into a multiwinner election~\cite{garg2021combatting}.

\subsection{Our Contribution}
In this paper, we take the standard view of redistricting as partitioning a planar graph on precincts into population-balanced, contiguous, and (in a heuristic sense) compact regions. We naturally extend the local fairness concept proposed in~\cite{agarwal2021locally} to this task. 

We first focus on the question of {\em auditing} a given plan for local fairness, that is, the non-existence of a population-balanced contiguous region in which a majority of voters are of the same party and is minority in the given plan. We show that this problem is computationally intractable in the worst case. Our first contribution is two heuristics for the auditing problem. Our first approach, that is scalable and practical, extends existing ensemble-based methods in a novel way: we assume the districts in the ensemble are the only districts to which voters can deviate, and given a plan to be audited, we test each of these districts as a potential deviation on that plan. Our second approach drills deeper into plans where the ensemble based method finds no deviating group; indeed, if the method found a deviating group, the plan was already deemed not locally fair. On the former set, we generate several random spanning trees, and devise a polynomial time dynamic programming algorithm that audits each tree for local fairness. If any of these audits finds a deviating group, the original plan was not locally fair. The dynamic program is not as efficient as the ensemble-based method; however, we provide empirical evidence that the ensemble method suffices to deem a plan locally fair, and the dynamic program typically does not find additional compact deviating groups. Finally, for redistricting plans that are not locally fair, we propose a measure that quantifies the unfairness of the plans by the portion of population with a justified complaint.

As our second contribution, we empirically study the notion of local fairness on real data on recent elections in the US. We generate plans using the (by now) standard \recom{}~\cite{deford2019recombination} ensemble method, and audit each plan for local fairness using the ensemble method, thereby producing an ordering of the plans via our unfairness measure. We empirically show that applying the criterion of local fairness prunes the space of candidate plans considerably, while still returning a set of potential candidates. Most global and statistical notions of fairness fail to do such pruning, since they are endogenously defined relative to the order statistics on the ensemble. We further show that not only is local fairness {\em achievable} on real redistricting tasks, but it is also compatible with extant global fairness properties. Indeed, when we compare locally fair plans and those with many deviating groups, the former tend to be just as compact, have comparable seat share outcomes, and sacrifice only a small amount of competitiveness. Thus local fairness can be used as an additional fairness criterion in conjunction with a global fairness criterion. We also investigate robustness of the local fairness concept, and show that fair redistricting plans remain consistent across different elections used. We finally show visualizations of fair and unfair plans; in particular showing that the visualization of deviating groups 
makes the local fairness notion explainable.

Taken together, our results demonstrate local fairness as an effective {\em pruning criterion} for candidate redistricting plans while sacrificing little in other desired properties. 
We also note that in practice, there could be other considerations when choosing the ``best'' plan even among many locally-fair plans; we leave the question of choosing these considerations to policy makers.


\section{Model and Preliminaries}
\label{s:model}
In keeping with recent literature~\cite{duchin2018discrete, deford2019recombination, cohen-addad2021computational, gurnee2021fairmandering}, the input to the redistricting problem is a planar connected graph $G = (V, E)$ where each vertex $v \in V$ represents an indivisible geographic unit (a precinct or a census block),\footnote{Typically, precincts are not split by redistricting plans~\cite{deford2019recombination}.} and an edge is placed between two vertices if they are geographically adjacent. Going forward, we refer to each $v \in V$ as a \emph{precinct} and $G$ as the \emph{precinct graph}.  



\noindent {\bf Redistricting Plans.}
For each precinct $v \in V$, let $\rho(v) > 0$ denote its population and let $\tau(v) \in [0, \rho(v)]$ denote the number of voters in $v$.\footnote{We assume that we know the exact number of people who cast a vote in each precinct, along with which candidate they voted for, such as is available for historical elections}. We let $\gamma(v) \in [0,1]$ and $\beta(v) = 1 - \gamma(v)$ denote the fraction of $\tau(v)$ who vote \emph{red} and \emph{blue}, respectively. 
Note that it is assumed each individual voter is exactly one of the two colors. 
For an arbitrary subset of precincts $W \subseteq V$, set $\rho(W) \coloneqq \sum_{v \in W} \rho(v)$, $\tau(W) \coloneqq \sum_{v \in W} \tau(v)$, $\gamma(W) \coloneqq \frac{\sum_{v \in W} (\gamma(v) \cdot \tau(v))}{\tau(W)}$, and $\beta(W) \coloneqq 1 - \gamma(W)$.  

\begin{definition}[$k$-redistricting plan]
A \emph{$k$-redistricting plan} of $G = (V,E)$ is a partition of $V$ into $k$ pairwise-disjoint subsets $D_1, D_2, \dots, D_k \subseteq V$, called \emph{districts}. Each district assumes the color of the majority of its voters. For a redistricting plan $\Pi$, let $B_\Pi$ (resp. $R_\Pi$) denote the set of precincts in blue (resp. red) districts in $\Pi$.
\end{definition}

In the following, we fix an error parameter $\eps > 0$,  and the desired number of partitions $k$. Note that the average population per district is $\frac{\rho(V)}{k}$.  We say a district $D \subseteq V$ is \emph{$\eps$-feasible} if: (1) $D$ induces a connected subgraph, and (2) the population of $D$ is at most $\eps$ away from average, {\em i.e.}, $(1-\eps) \cdot \frac{\rho(V)}{k} \leq \rho(D) \leq (1+\eps) \cdot \frac{\rho(V)}{k}$.  A redistricting plan $\Pi$ is \emph{$\eps$-feasible} if each district $D_i \in \Pi$ is $\eps$-feasible.  

We note that this definition of an $\eps$-feasible plan is consistent with the general practice in the U.S, where the sizes of districts should be balanced in terms of their population, based on census information, not in terms of the number of eligible voters. 
Since $\eps$ and $k$ will be fixed throughout, we drop the prefixes and refer to $k$-redistritcting plans and $\eps$-feasible districts as \emph{redistricting plans} and \emph{feasible} districts, respectively.

 \noindent {\bf Local Fairness.}
We extend the notion of \emph{local fairness} proposed by~\cite{agarwal2021locally} to the graph-based redistricting problem.  We say that a feasible district $W \subseteq V$ is \emph{red-majority} (\emph{red} in short) if $\gamma(W) \geq \beta(W)$, and \emph{blue-majority} (blue) otherwise. We call this majority color as the {\em color} of $W$. Given a redistricting plan $\Pi$, any voter whose color agrees with the color of its assigned district in $\Pi$ is deemed {\em happy} with respect to $\Pi$, and the remaining voters are {\em unhappy}. 

\begin{definition}[$c$-locally fair] \label{def:dg}
Given a feasible redistricting plan $\Pi$ of $G$ and a constant $c \in [1/2, 1]$, a feasible district $W \subseteq V$ is a \emph{red $c$-deviating group} with respect to $\Pi$ if $W$ is red and at least a $c$-fraction of its voters are unhappy red voters in $\Pi$, or formally,  $\sum_{v \in W \cap B_\Pi} \gamma(v) \cdot \tau(v) > c \cdot \tau(W).$
A \emph{blue $c$-deviating group} is defined analogously. 
We call a feasible redistricting plan $\Pi$ of $G$ \emph{$c$-locally fair} if there are no red or blue $c$-deviating groups with respect to $\Pi$.
\end{definition}

When $c = 1/2$, only a simple majority of voters in a deviating group must be unhappy. In this special case, we omit the prefix $c$ by referring to red deviating groups, blue deviating groups, and locally fair redistricting plans. Throughout, we refer to locally fair redistricting plans as fair plans. 

We are thus interested in the following two problems.

\begin{description}
\item[\audit\ problem.]
Given a feasible redistricting plan $\Pi$ and a parameter $c \in [1/2, 1]$, decide whether $\Pi$ is $c$-locally fair.
\item[\generation\ problem.] Given a precinct graph $G$ and parameters $\eps, k$ and $c$, compute a feasible redistricting plan $\Pi$ of $G$ such that $\Pi$ is $c$-locally fair, or report that none exists.
\end{description}

For a redistricting plan $\Pi$ that is not locally fair, we quantify its degree of unfairness as follows. Consider all deviating groups of $\Pi$, and define the unfairness score of $\Pi$ as the fraction of all voters that are unhappy in some deviating group. 
Formally, let 
\[ W^\ast_{\text{B}}(\Pi) \coloneqq R_\Pi \cap \left( \bigcup \big\{ W \subseteq V \mid W \text{ is a blue deviating group of $\Pi$} \big\} \right) \]
denote the set of red precincts that lie in some blue deviating group of $\Pi$. Similarly, define $W^\ast_{\text{R}}(\Pi)$. Then the unfairness score of $\Pi$ is defined as: 

\begin{small}
\[ \unf(\Pi) \coloneqq \frac{\beta(W^\ast_{\text{B}}(\Pi)) \cdot \tau(W^\ast_{\text{B}}(\Pi)) + \gamma(W^\ast_{\text{R}}(\Pi)) \cdot \tau(W^\ast_{\text{R}}(\Pi))}{\tau(V)}. \]
\end{small}

This score captures the fraction of voters that are both (i) unhappy in $\Pi$ and (ii) in the majority color of some deviating group of $\Pi$. Note that $\unf(\Pi) \in [0,1]$, and equals zero if $\Pi$ is locally fair.

\noindent {\bf Compactness.} In addition to requiring that districts be contiguous and population balanced, many  redistricting models also require that the districts be \emph{compact}. 
However, in contrast to the former two criteria, there is no universally agreed measure of compactness~\cite{becker2020redistricting, gurnee2021fairmandering}. 
For example, the Princeton Redistricting Report Cards\footnote{\url{https://gerrymander.princeton.edu/}} uses Reock~\cite{reock1961note} and Polsby-Popper~\cite{polsby1991third} scores, both of which are derived from the area and perimeter of the geographic districts drawn. It also uses the number of counties split into multiple districts. In the discrete model of precinct graphs, one common measure of compactness is the number of cut edges formed by the plan, i.e., the number of edges whose endpoints lie in different districts of $\Pi$~\cite{deford2019recombination, cohen-addad2021computational}. Though we do not enforce compactness in the generation and audit problems, the algorithms we use are biased towards compact plans, as we empirically demonstrate.


\noindent{\bf Organization.}
In Section~\ref{s:algorithm}, we present two algorithms for the \audit\ and \generation\ problems. We then describe the experimental setup and empirical results in Section~\ref{s:exp}. 
Additional methodological and experimental results are presented in the Supplementary Material.
\section{Heuristics for Efficiently Auditing Local Fairness}
\label{s:algorithm}
In Appendix~\ref{ss:hardness}, we prove that both the \audit\ and \generation\ problem are \textsf{NP}-complete, thereby necessitating heuristic or approximately optimal approaches.

Our main contribution in this section is efficient heuristics for \audit. The methods trade off computational efficiency for accuracy in determining local fairness. We describe the types of inaccuracy (one-sided versus two-sided error) that arise after describing the methods. We also provide empirical evidence that the more computationally efficient of these methods suffices to deem plans as locally fair. For \generation, we simply generate an ensemble of feasible redistricting plans using existing MCMC approaches, and run the \audit\ algorithm to find the unfairness score $\unf(\Pi)$ for each generated plan $\Pi$, thereby ranking the plans in the ensemble in terms of fairness.

\subsection{Ensemble-based Auditing}
\label{ss:ensemble}
The computationally more efficient approach makes use of ensembles in a novel way as follows:

\begin{enumerate}
    \item {\bf Generation.} Generate an ensemble of $t$ redistricting plans $\ens = \{\Pi_1, \Pi_2, \ldots, \Pi_t\}$.
    \item {\bf Districts to test.} Let $\Delta = \bigcup_{j =1}^t \Pi_j$
    be the collection of districts in the plans in $\ens$. 
    \item {\bf Audit.} Treat $\Delta$ as the candidate set of deviating groups. For each plan $\Pi \in \ens$, compute if each $D \in \Delta$ is a $c$-deviating group (either red or blue). This step yields, for each $\Pi \in \ens$, the set $\Tilde{W}^\ast_{\text{B}}(\Pi) \subseteq V$ of precincts in blue deviating groups $D \in \Delta$ (which we use as an approximation of $W^\ast_{\text{B}}(\Pi)$), and similarly $\Tilde{W}^\ast_{\text{R}}(\Pi)$. We then use  $\Tilde{W}^\ast_{\text{B}}(\Pi)$ and  $\Tilde{W}^\ast_{\text{R}}(\Pi)$ to compute the unfairness score $\unf(\Pi)$ for each $\Pi \in \ens$.
    
\end{enumerate}


For the first step of an ensemble of plans $\ens$, we use the \recom\ algorithm~\cite{deford2019recombination}. This algorithm develops a Markov Chain whose state space is the full set of ($\eps$-)feasible ($k$-)partitions of $G$. In each step, it randomly combines two (or more in its general form) districts of the current redistricting plan, generates a random spanning tree of the subgraph induced by the combined districts, and re-partitions the subgraph into two parts by cutting edges in the spanning tree so that the new resulting districts remain population balanced. The random spanning tree step \emph{biases} the Markov chain towards plans with compact districts, where compactness is measured by number of cut edges~\cite{procaccia2022compact}.   

In the remainder of the paper, we describe our auditing framework when \recom\ is used as the ensemble generation technique. We note however that we can use any other method that produces compact plans, for instance, iterative-merging flood-fill algorithms~\cite{chen2013unintentional}.

\subsection{Auditing via Dynamic Programming on Trees} 
\label{ss:treedp}
The ensemble-based \audit\ method makes one-sided error -- if it finds a deviating group, then the plan is not fair, while the absence of a deviating group from $\Delta$ only means the plan is {\em likely} fair. Indeed, in our experiments, there are often multiple redistricting plans that the ensemble-based audit deems likely fair, calling for a more systematic method for further analyzing the candidate plans that are potentially fair. This motivates the approach we now describe. 

We show that the \audit\ problem is efficiently solvable if $G$ is a tree. We then exploit this fact and use this as a heuristic to solve \audit\ on a general planar graph $G$, given partition $\Pi$, as follows: 

\begin{enumerate}
    \item {\bf Random spanning trees.} First generate a collection of random spanning trees $\mathcal{T}$ of $G$.
    \item {\bf Audit each tree.} For each tree $T \in \mathcal{T}$, decide if there exists a feasible deviating group with respect to $\Pi$ such that this district is a connected subtree of $T$. Note that such a district will also be a connected subgraph in $G$.
\end{enumerate}


We implement step $2$ for a tree $T$ using dynamic programming. 
We check for the existence of a blue deviating group for $c = 1/2$ in a tree $T$ as follows; the case for red groups and larger $c$ follows similarly. 
Note that the range for the size of a district is $[(1-\eps)\sigma, (1+\eps)\sigma]$, where $\sigma = \rho(V)/k$. For each precinct $v \in V$, let $D(v)$ denote the district that contains $v$ in $\Pi$. Let the number of unhappy blue individuals in precinct $v$ be denoted $\text{uhp}(v) = \beta(v) \cdot \tau(v)$ if $D(v) \in R_\Pi$, and zero otherwise. The task is to decide whether there is a connected subtree $W \subset T$ such that $\sum_{v \in W} \rho(v) \in [(1-\eps)\sigma, (1+\eps)\sigma]$ and $\sum_{v \in W} \text{uhp}(v) > \frac{1}{2} \cdot \sum_{v \in W} \tau(v)$.

For each $v \in V$, let $T_v$ be the subtree of $T$ rooted at $v$. For a vertex $v \in V$ and two parameters $i,p \leq (1+\eps)\sigma$, let $A[v,i,p]$ denote the maximum number of unhappy blue voters in a subtree $W \subseteq T_v$ such that $v \in W$, $\tau(W) \leq i$, and $\rho(W) = p$. To compute $A[v, i, p]$, we take the best subset of children of $v$ such that unhappy blue voters are maximized subject to population and voter constraints. 

For a leaf $v$ of $T$, $A[v,i,p] = \text{uhp}(v)$ if $i \geq \tau(v)$ and $p = \rho(v)$, and $A[v,i,p] = 0$ otherwise. 

Next, let $v$ be an internal node of $T$. 
Let $U = \text{children}(v) = \{u_1, \ldots, u_{\text{deg}(v)}\}$ denote the set of $v$'s children in $T$. Let $i' = i - \tau(v)$ and $p' = p - \rho(v)$. We have
\begin{align}
    A[v,i,p] = \text{uhp}(v) + \max\limits_{\{i_j, p_j\}} \bigg\{ \sum_{j=1}^{\text{deg}(v)} A[u_j, i_j, p_j] \mid \bigg({\sum_{j=1}^{\text{deg}(v)} i_j \leq i'}\bigg) \land \bigg({\sum_{j=1}^{\text{deg}(v)} p_j = p'}\bigg) \bigg\}. \label{vertical_dp}
\end{align}

To compute the second term in Equation~(\ref{vertical_dp}) efficiently, let $B_{v,i,p}[j,x,y]$ denote the maximum number of total unhappy blue voters in a union of subtrees rooted at $\{u_1, \ldots, u_j\}$ with total voter count at most $x$ and total population level $y$. The second term of the RHS in Equation~(\ref{vertical_dp}) is $B_{v,i,p}[\text{deg}(v),i',p']$.
We then have
$B_{v,i,p}[1,x,y] = A[u_1,x,y]$ for all $x,y$, and for all $j \geq 2$ we have
\begin{align}
    B_{v,i,p}[j,x,y] = \max_{x' \in [0,x], y' \in [0,y]} \left\{ B_{v,i,p}[j-1,x-x',y-y'] + A[u_j, x', y'] \right\}. \label{horizontal_dp}
\end{align}

The algorithm thus proceeds by computing all $A[v,i,p]$ in an outer loop in increasing order of $i$ and $p$ and in bottom-up order of $v$. Each computation of $A[v,i,p]$ proceeds with an inner loop that computes $B_{v,i,p}[j,x,y]$ in increasing order of $j$, $x$, and $y$. For each $v$, after computing all $A[v,i,p]$, we check if there is any $p \in [(1-\eps)\sigma, (1+\eps)\sigma]$ such that $A[v,i,p] > \frac{i}{2}$. If so, there exists a blue deviating group rooted at $v$ with population $p$, total voter count at most $i$, and total number of unhappy blue voters $A[v,i,p]$. 

\noindent {\bf Time complexity.} The algorithm computes $O(\abs{V} \cdot \sigma^2)$ values of $A[v,i,p]$, each computing $O(\mbox{deg}(T) \cdot \sigma^2)$ values of $B_{v,i,p}[j,x,y]$. 
Computing each $B_{v,i,p}[j,x,y]$ requires $O(\sigma^2)$-time to loop through all values of $x'$ and $y'$. Hence the overall time complexity is $O(\abs{V} \cdot \sigma^6 \cdot \mbox{deg}(T))$.  
We conclude the following.

\begin{theorem}
\label{thm:dp1}
\audit\ problem on a tree $T(V,E)$ can be solved in time $O(\abs{V} \cdot \sigma^6 \cdot \mbox{deg}(T))$. Here, $\mbox{deg}(T)$ is the maximum degree of a node in $T$.
\end{theorem}

\paragraph{Ensemble-based Auditing is Approximately Sufficient}  
The dynamic programming method is less desirable because of two significant limitations. The first is its running time, which is prohibitively high for practical use. The second limitation is the introduction of a different type of error from the ensemble-based method. Note that as with the ensemble-based method, the non-existence of a deviating group in the DP only provides high confidence (but not absolute certainty) about the plan being locally fair. At the same time, there is a more subtle type of error the DP can make in the other direction: even if it finds a deviating group on the tree, this may not be a {\em compact} deviating group in the original graph. 
Therefore, the algorithm can output ``not locally fair" when the only deviating groups are non-compact. We therefore need a final step where we check the deviating group to see that it is both feasible and compact on the original graph.

In Appendix~\ref{ss:dp_details}, we first show approaches to speed up the running time of the dynamic program to make it practical. In Appendix~\ref{ss:exp_dp}, we describe additional assumptions under which the dynamic program runs efficiently, we empirically show that it fails to find compact $c'$-deviating groups in plans that the ensemble-based audit method deems $c$-locally fair, for $c'$ slightly larger than $c$. On the other hand, it easily finds such deviating groups in plans that the ensemble-based method deems unfair. In summary, the dynamic programming approach obviates its own use in practice, since we provide strong evidence that the ensemble-based auditing method suffices in order to deem redistricting plans as locally fair, assuming the strength $c$ of the deviating group is relaxed slightly.


\section{Experiments}
\label{s:exp}
In our experiments, we attempt to answer the following questions. First, is local fairness achievable on real redistricting tasks? Second, is it compatible with extant measures of global fairness? Finally, is the notion robust if the underlying data changes? Given the previous discussion, our experiments in this section focus exclusively on the ensemble based method. 
In Appendix~\ref{s:dp_exp} we discuss the experiments for auditing via dynamic programming.

\subsection{Datasets and Methods} 
\label{ss:set-up}
All data used in our experiments is obtained from the MGGG States open repository~\cite{mgggstates}. We obtain shapefiles and precinct graphs for Massachusetts (MA), Maryland (MD), Michigan (MI), North Carolina (NC), Pennsylvania (PA), Texas (TX), and Wisconsin (WI).\footnote{These are chosen to represent a spectrum from states whose elections are typically competitive (e.g., NC and WI) to states whose elections are typically lopsided (e.g., MA and TX).} We set $\rho(v)$ to the 2010 census total population in each precinct $v$. The default election we use is the 2016 presidential election, while the 2012 presidential election is also used in the robustness tests.  For each precinct $v$, we collect the number of total votes for the Republican party $r_v$ and the total vote amount for the Democrat party $b_v$ in the 2016 presidential election from~\cite{mgggstates}. We set $\gamma(v) = r_v/(r_v+b_v)$, $\beta(v) = 1-r_v$, and $\tau(v) = r_v + b_v$, so $\tau(v)$ is the total number of red (Republican) and blue (Democrat) voters. 

Using the \recom\ algorithm, we generate an ensemble of redistricting plans for each state.\footnote{Note that \recom\ generates plans without taking into account electoral data.} Each ensemble consists of $1,000$ redistricting plans, each being the outcome of an independent $10,000$-step Markov chain (with default population balance parameter $\eps = 0.02$) seeded with a recent congressional electoral plan of the state. We set $k$ to be the number of congressional districts in the 2016 election in each state. We then obtain the collection $\Delta$ of candidate districts for each state by taking the union of the districts in each plan in the ensemble.  The properties of input graphs of states and their corresponding ensembles are summarized in Table~\ref{tab:inputs}. 

\begin{table}[!t]

 \parbox{\linewidth}{   
    \centering
    \begin{tabular}{||c c c c c c c||} 
     \hline
     State & $|V|$ & $|E|$ & $\rho(V)$ & $\tau(V)$ & $k$ & $\abs{\Delta}$ \\ \hline \hline
     MA    & 2.1K  & 5.9K  & 6.55M  & 5.13M  & 9   & 8.5K \\ \hline
     MD    & 1.8K  & 4.7K  & 5.77M  & 4.42M  & 8   & 7.8K \\ \hline
     MI    & 4.8K  & 12.5K & 9.88M  & 7.54M  & 14  & 14.0K \\ \hline
     NC    & 2.7K  & 7.6K  & 9.53M  & 7.25M  & 13  & 12.9K \\ \hline
     TX    & 8.9K  & 24.7K & 25.14M & 18.28M & 36  & 35.9K \\ \hline
     PA    & 9.2K  & 25.7K & 12.70M & 9.91M  & 18  & 18.0K \\ \hline
     WI    & 7.1K  & 19.5K & 5.69M  & 4.35M  & 8   & 7.87K \\ \hline
    \end{tabular} 
    \caption{Properties of data and ensembles.}
    \label{tab:inputs}}
  \end{table}


\begin{table}[!thbp]
 \parbox{\linewidth}{   
        \centering
    \begin{tabular}{||c | c c c c c c c||} 
    \hline
    State & MA & MD & MI & NC & PA & TX & WI \\ 
    \hline \hline
    $c = .5$ & 100\% & 26.7\% & 0.3\% & 4.9\% & 12.0\% & 2.8\% & 76.9\% \\
    $c = .51$ & 100\% & 27.0\% & 1.6\% & 8.7\% & 28.6\% & 5.4\% & 83.6\% \\
    $c = .52$ & 100\% & 28.0\% & 11.9\% & 14.8\% & 41.4\% & 8.7\% & 87.2\% \\
    $c = .55$ & 100\% & 31.5\% & 35.2\% & 63.9\% & 95.2\% & 26.0\% & 94.0\% \\
    \hline
    \end{tabular}
    \caption{Percent of plans without $c$-deviating groups for different $c$ values.}
    \label{tab:achievability}}
\end{table}

\subsection{Locally Fair Plans: Counts and Visualizations}
\label{ss:achievability}

\begin{figure}[!htbp]
    \begin{subfigure}[t]{0.225\columnwidth}
        \centering
        \includegraphics[width=\columnwidth]{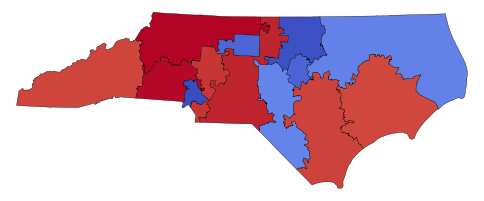}
        \caption{NC Fair Plan}
        \label{fig:nc_fair_map}
    \end{subfigure}
    \hfill
        \begin{subfigure}[t]{0.225\columnwidth}
        \centering
        \includegraphics[width=\columnwidth]{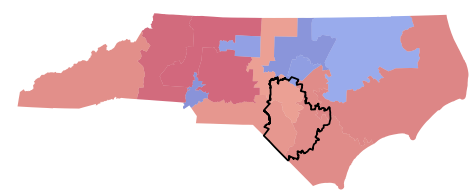}
        \caption{NC Plan with blue dev. group}
        \label{fig:nc_dg_example}
    \end{subfigure}
    \hfill
    \begin{subfigure}[t]{0.215\columnwidth}
    \centering
    \includegraphics[width=\columnwidth]{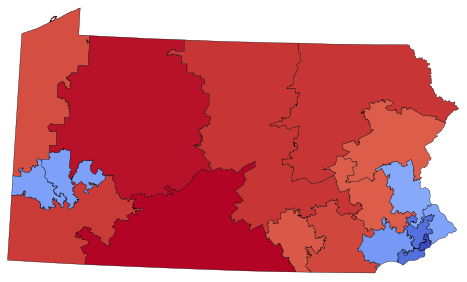}
    \caption{PA Fair Plan}
    \label{fig:pa_fair_map}
    \end{subfigure}
    \hfill
    \begin{subfigure}[t]{0.215\columnwidth}
    \centering
    \includegraphics[width=\columnwidth]{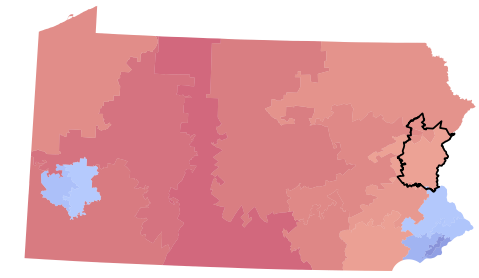}
    \caption{PA Plan with blue dev. group}
    \label{fig:pa_dg_example}
    \end{subfigure}
    \caption{North Carolina and Pennsylvania plans without and with deviating groups. The blue deviating groups are drawn with a black outline. In these figures, the districts are coded by its color and the extent of partisanship: districts with a larger value of $\gamma$ (resp. $\beta$) are colored in darker red (resp. blue).}
    \label{fig:plans}
\end{figure}

We first use the ensemble-based audit method to audit each plan in the ensemble against the set $\Delta$ of districts from the entire ensemble. For each $c \in \{0.5, 0.51, 0.52, 0.55 \}$, we count the number of plans in the ensemble without $c$-deviating groups in $\Delta$. We present the results in Table~\ref{tab:achievability}. For all values of $c$, there exists plans in the ensemble that admit no $c$-deviating groups in $\Delta$ and thus are identified locally fair by the ensemble-based algorithm. Clearly, a larger value of $c$ implies more plans are identified as locally fair. With $c = 1/2$, very few (but a non-zero number of) plans are identified as fair in four of the seven states. Every plan in the MA ensemble is identified as fair with all districts won by one party. Hence, we omit MA from all subsequent experiments.  

In Figures~\ref{fig:nc_fair_map} and~\ref{fig:pa_fair_map}, we present examples of fair plans (with no $0.5$-deviating groups) in NC and PA respectively, while in Figures~\ref{fig:nc_dg_example} and~\ref{fig:pa_dg_example}, we present  ``unfair'' plans with many $0.5$-deviating groups. 
We show visualizations for other states in Appendix~\ref{ss:visualization}.

\begin{figure}[!htbp]
    \begin{subfigure}[b]{0.16\columnwidth}
    \centering
    \includegraphics[width=\columnwidth]{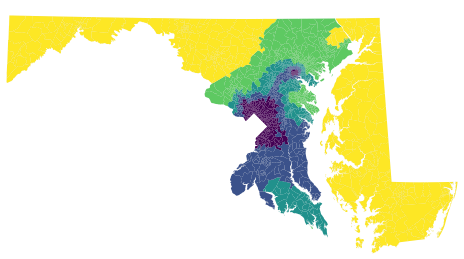}
    \caption{MD}
    \label{fig:md_precincts}
    \end{subfigure}
    \hfill
    \begin{subfigure}[b]{0.125\columnwidth}
    \centering
    \includegraphics[width=\columnwidth]{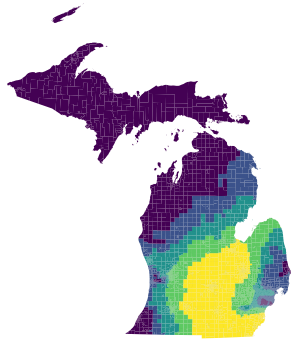}
    \caption{MI}
    \label{fig:mi_precincts}
    \end{subfigure}
    \hfill
    \begin{subfigure}[b]{0.195\columnwidth}
    \centering
    \includegraphics[width=\columnwidth]{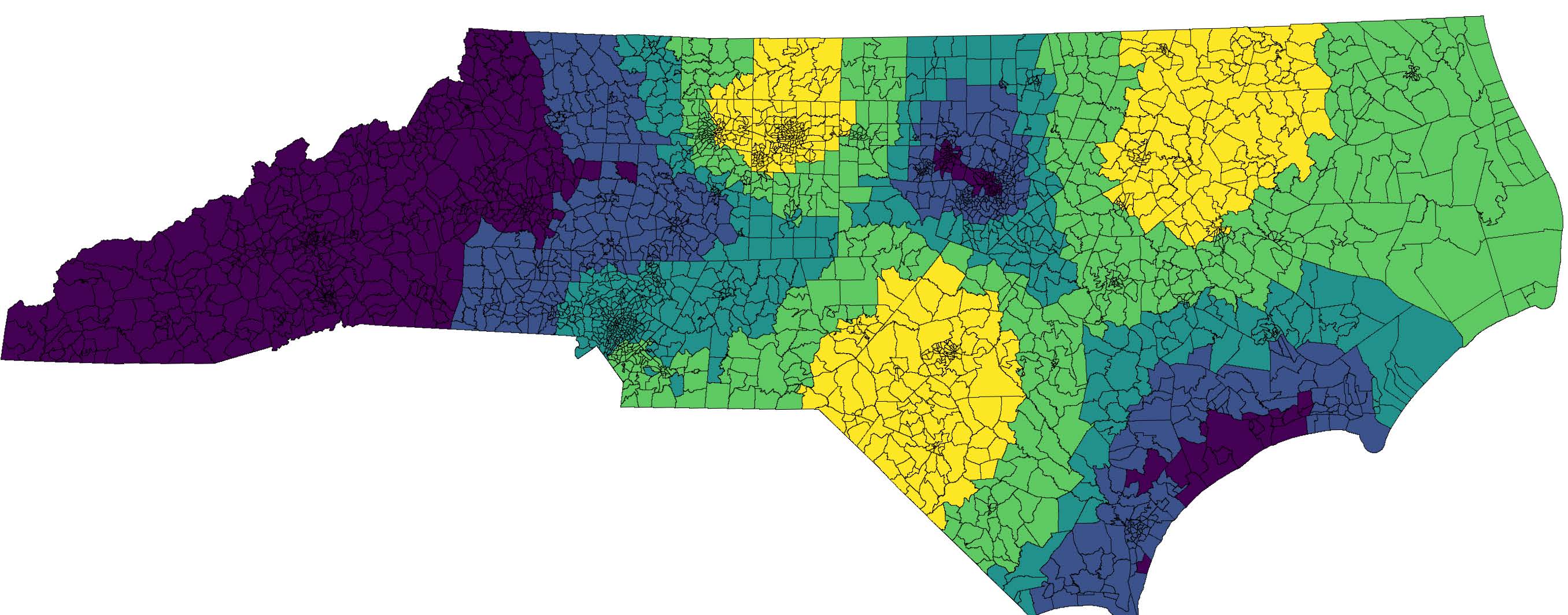}
    \caption{NC}
    \label{fig:nc_precincts}
    \end{subfigure}
    \hfill
    \begin{subfigure}[b]{0.15\columnwidth}
    \centering
    \includegraphics[width=\columnwidth]{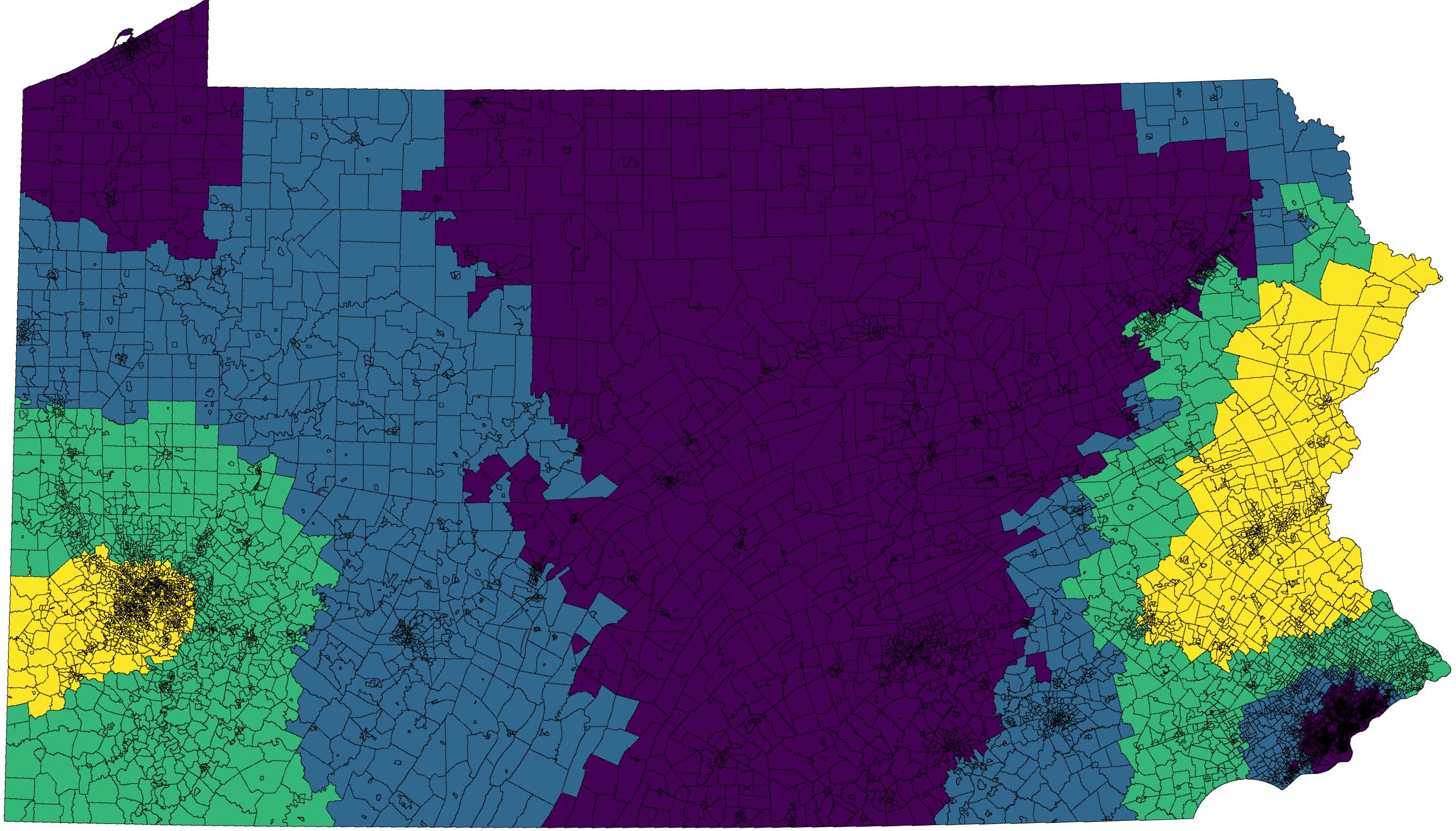}
    \caption{PA}
    \label{fig:pa_precincts}
    \end{subfigure}
    \hfill
    \begin{subfigure}[b]{0.16\columnwidth}
    \centering
    \includegraphics[width=\columnwidth]{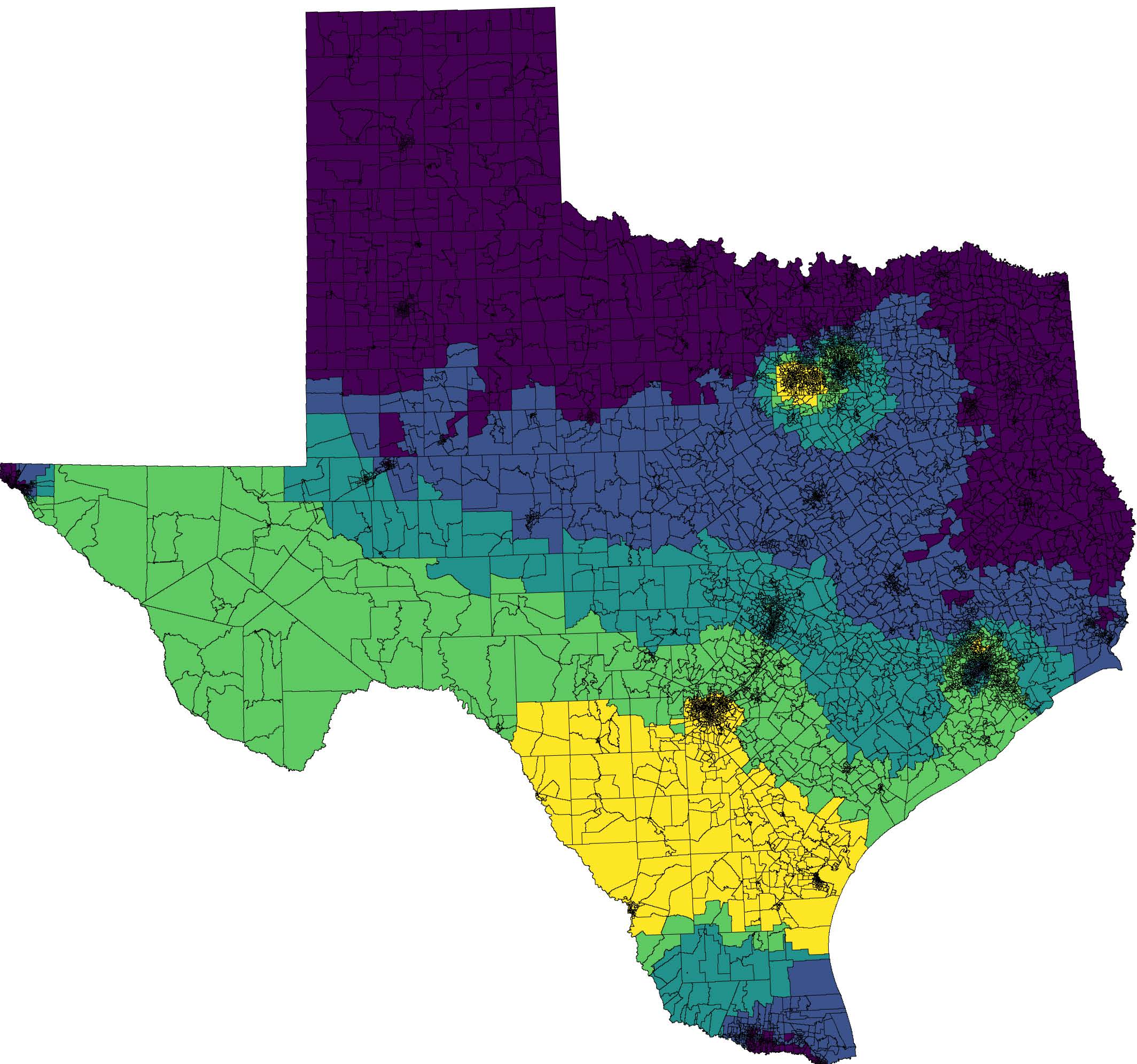}
    \caption{TX}
    \label{fig:tx_precincts}
    \end{subfigure}
    \hfill
    \begin{subfigure}[b]{0.15\columnwidth}
    \centering
    \includegraphics[width=\columnwidth]{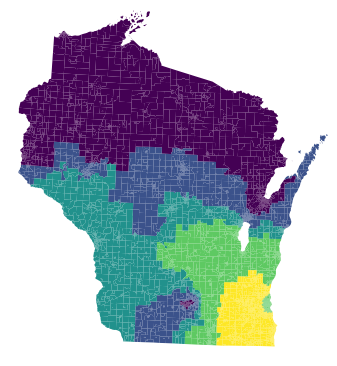}
    \caption{WI}
    \label{fig:wi_precincts}
    \end{subfigure}
    \caption{Precincts shaded in lighter color are contained in more deviating groups.}
    \label{fig:precincts}
\end{figure}

We note that a large fraction of plans are locally fair in some states (WI), while others have many deviating groups, even for a large setting of $c$ (MI). To understand where deviating groups are located, in Figure~\ref{fig:precincts}, we plot a heat-map of the precincts, counting the number of $1/2$-deviating groups (of either color) that contain that precinct, with yellow representing large counts and purple representing low counts. 
In every state except MD,\footnote{We discuss MD in more detail in Appendix~\ref{ss:visualization}.} we observe that precincts with highest counts are located either in an urban area or in proximity of one. This phenomenon is consistent with the perceived correlation between voter distribution and type of residence~\cite{borodin2018big, wheeler2020impact}. In states with multiple dense urban areas (e.g., NC), there is sufficient flexibility in the redistricting process to ``crack'' a highly-concentrated urban demographic into multiple districts. In this case, the urban area may form a deviating group resulting in high numbers of unfair plans.

We note that our visualizations -- in particular, the deviating groups in unfair plans, as well as the heat-map of likelihood of unhappiness of a precinct -- make the local fairness concept {\em explainable}.

\begin{figure}[!htbp]
\centering
    \begin{subfigure}[b]{0.98\textwidth}
    \centering
    \includegraphics[width=\textwidth]{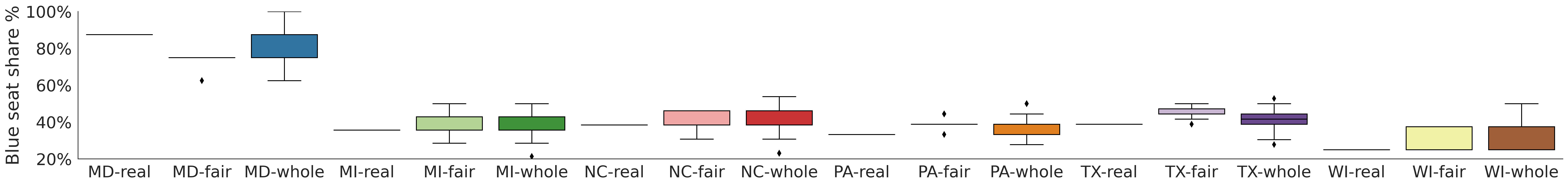}
    \caption{Seat shares as percent of districts with blue majority.}
    \label{fig:seat_distribution}
    \end{subfigure}
    \hfill
    \begin{subfigure}[b]{0.98\textwidth}
    \centering
    \includegraphics[width=\textwidth]{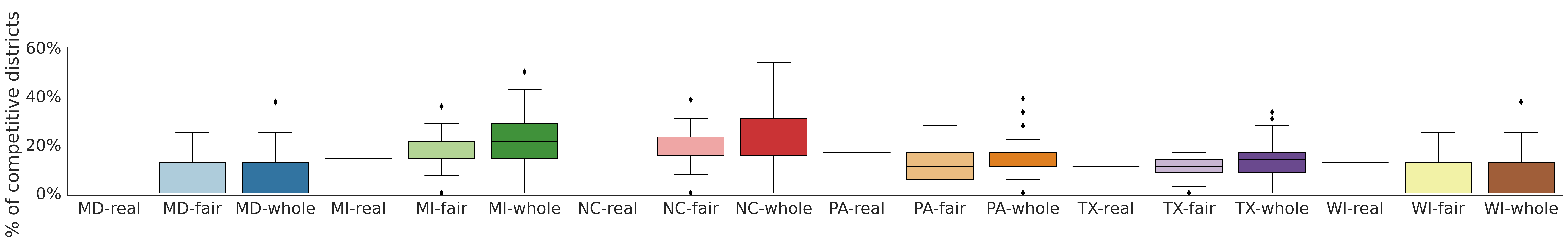}
    \caption{Percentage of districts being competitive (having less than $53.5\%$ voters in majority).}
    \label{fig:competitiveness}
    \end{subfigure}
    \begin{subfigure}[b]{0.98\textwidth}
    \centering
    \includegraphics[width=\textwidth]{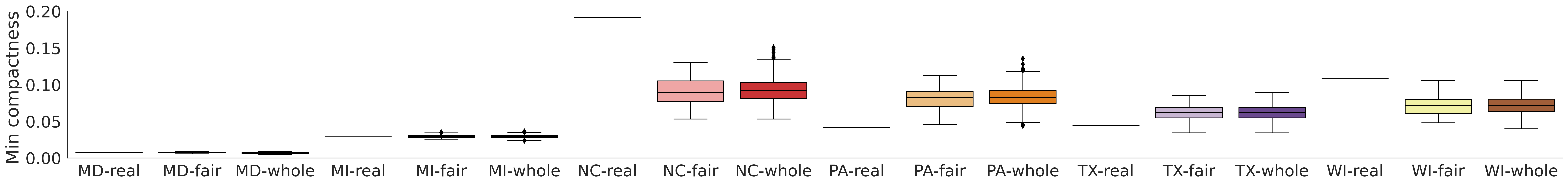}
    \caption{Minimum Polsby-Popper score of districts in a plan.}
    \label{fig:compactness_min}
    \end{subfigure}
\caption{Distribution of fairness and compactness metrics among subsets of generated plans}
\end{figure}

\subsection{Compatibility with Extant Fairness and Compactness Notions}
\label{ss:compatibility}
The ensemble-based auditing approach can be viewed as a pruning method that identifies a subset of plans as locally fair. We now ask: how do locally fair plans  perform on extant global fairness and compactness criteria compared to average plans in the ensemble? Towards this end, for $c = 1/2$, we rank the plans in the ensemble by the unfairness score $\unf(\Pi)$. We compare properties of the top 5\% plans in the ranking (which are most locally fair) against the entire ensemble, as well as a recent enacted congressional redistricting plan.\footnote{If more than the $5\%$ of the plans are locally fair, we take an arbitrary subset of the locally fair plans to serve as the top $5\%$.} 

\noindent {\bf Seat share outcomes.} For each plan, we compute \emph{Blue\%}, the percentage of seat shares claimed by the Democratic candidate.
The resulting distribution is shown in Figure~\ref{fig:seat_distribution}. The seat share distributions of the top 5\% locally fair plans are comparable to the entire ensemble, sometimes achieving lower variance.

\noindent {\bf Number of competitive districts.}
The Princeton Gerrymandering Project~\cite{princeton} defines a district as \emph{competitive} if the majority color is at most $53.5$\% of the total votes. 
Following this definition, in Figure~\ref{fig:competitiveness}, we compare the percent of competitive districts in a plan. The locally fair plans produce slightly fewer competitive districts: since larger majorities reduce the number of unhappy voters, finding a deviating group becomes harder. However, there exist fair plans that produce the median percentage of competitive districts in the ensemble in all but one state, and they are comparable with or better than the enacted plan in all states. We present results using a different competitiveness metric in Appendix~\ref{ss:alternative_exps}.

\noindent {\bf Minimum compactness.} Another measure of quality of districting plans is compactness -- a non-compact district not only makes less geographic sense, but is also more likely to have been gerrymandered to favor one party over another. Two commonly used metrics to evaluate the compactness in redistricting plans are the average and minimum Polsby-Popper scores~\cite{polsby1991third} of the districts in the plan~\cite{princeton}: For a district $D \in \Pi$, the Polsby-Popper score is defined as $4\pi A(D)/P(D)^2,$ where $A(D)$ and $P(D)$ are the area and the perimeter of the planar region $D$, respectively; a higher value implies a more compact district. In Figure~\ref{fig:compactness_min}, we show that the minimum compactness of the locally fair plans remains comparable to that of the entire ensemble. We present results on the average Polsby-Popper score in Appendix~\ref{ss:alternative_exps}. 

Taken together, our results show that local fairness is compatible with fair seat share and compactness, while sacrificing only a small amount on number of competitive districts.

\subsection{Actual Plans and Additional Results}

We also compute the local fairness of plans actually enacted for previous elections. As it is relatively easy to find a locally fair plan in WI, its actual plan is indeed locally fair, while the actual plans for all other states are not locally fair.\footnote{Note that all experiments use the 2016 presidential election data, while the plans in use are mostly drawn in 2011, except the NC one that is drawn in 2019.} Using the $\unf$ ranks, the enacted plan falls in the $55^{th}$ percentile (fairer than $45\%$ of the plans) in MD, $73^{th}$ percentile in MI, $84^{th}$ percentile in NC, $29^{th}$ percentile in PA, and $22^{nd}$ percentile in TX. While the MI and NC plans are (somewhat surprisingly) above average in local fairness, they have very few (or no) competitive districts. In general, enacted plans that achieve above average local fairness compared to the ensemble (MD, MI, and NC) have fewer competitive districts, and enacted plans with more competitive districts perform below average in local fairness. On the other hand, our results demonstrate that it is possible to find locally fair plans with a comparable (to the ensemble) amount of districts remaining competitive.

In Appendix~\ref{ss:robustness}, we show that the locally fair notion is also robust to voting patterns in the sense that there is a large overlap between fair plans identified by auditing the ensemble using the voter data from the 2012 and the 2016 presidential elections for the states of NC, PA, and TX. 






\section{Conclusions}
In summary, in contrast to extant global notions of fairness in redistricting such as seat distribution or district competitiveness, the notion of local fairness is an explainable notion that mitigates justified complaints of populations in compact geographic regions. Our experiments show that local fairness is not only possible to achieve in practice, but choosing locally fair plans also does not come at the expense of other global fairness properties. 

Several open questions arise from our work. In terms of algorithm design, it is an open question of whether there is an approximation algorithm for either the \audit\ or \generation\ problem, and whether such an algorithm could take into account compactness. It would also be interesting to extend our methods to capture additional real-world criteria used in redistricting, such as a penalty for splitting up counties, or a requirement for a majority-minority district. In particular, can fair plans be locally modified so that they remain fair and such real-world criteria are satisfied? 

Finally, our exploration of robustness of local fairness to voter turnout is preliminary, since it compares the outcomes between two election data in one state. It would be interesting to extend our work to a stochastic setting, where each individual in the population has a ``likely voter'' score (or probability to vote), and we need high confidence in the non-existence of a $c$-deviating group. 


\newpage

\bibliographystyle{abbrv}
\bibliography{ref.bib}

\newpage
\appendix
\label{s:apdx}
\centerline {\LARGE \bf  Appendix}

\section{NP-Hardness Proofs} 
\label{ss:hardness}

\begin{theorem} \label{thm:audit_hardness}
\audit\ is \textsf{NP}-complete.
\end{theorem}

\begin{proof}
We reduce the \textsf{NP}-complete Connected $k$-Subgraph Problem on Planar Graphs with Binary Weights (\ckspb)~\cite{hochbaum1994node} to the \audit\ problem. 
Given a connected planar graph $\mathsf{G} = (\mathsf{V}, \mathsf{E})$ where $\mathsf{V} = \{\mathsf{v}_1, \mathsf{v}_2, \ldots, \mathsf{v}_{\abs{\mathsf{V}}}\}$, a vertex weight $\omega(\mathsf{v}_i) \in \{0,1\}$ for each $\mathsf{v}_i \in \mathsf{V}$, a size $M \geq 2$, and a targeted total weight $\Omega \in \mathbb{Z}^{+}$,\footnote{We use $M$ and $\Omega$ here to avoid confusing with the notations $k$ and $W$ in our problem.} the decision version of \ckspb\ asks whether there is a subset $\mathsf{W} \subseteq \mathsf{V}$ of $M$ vertices such that its induced subgraph $\mathsf{H} \subseteq \mathsf{G}$ is connected and $\sum_{\mathsf{v}_i \in \mathsf{W}} \omega(\mathsf{v}_i) \geq \Omega$. Here it is assumed that $M \leq \abs{\mathsf{V}}$ and $\omega(\mathsf{v}_i) < \Omega$ for each $\mathsf{v}_i \in \mathsf{V}$; otherwise the problem is trivial. 

Given an arbitrary instance of \ckspb, we construct an instance for \audit\ as follows. For each vertex $\mathsf{v}_i \in \mathsf{V}$, we construct two precinct nodes $v_i$ and $u_i$, and an edge between $v_i$ and $u_i$. For each edge $(\mathsf{v}_i, \mathsf{v}_j) \in \mathsf{E}$, we construct an edge between $v_i$ and $v_j$. Hence, the resulting graph $G = (V,E)$ has $2\abs{\mathsf{V}}$ precinct nodes and $\Paren{\abs{\mathsf{V}}+\abs{\mathsf{E}}}$ edges, and is still planar. 

For all $i = 1, 2, \ldots, \abs{\mathsf{V}}$, we let $\rho(v_i) = \tau(v_i) = 2M \Omega$ and $\gamma(v_i) = \Paren{\frac{M-1}{2M}+\frac{\omega(\mathsf{v}_i)}{2\Omega}}$; note that each $\gamma(v_i)$ is in $(0, 1)$ and each $\gamma(v_i) \cdot \tau(v_i)$ is an integer. For all $i$, we let $\rho(u_i) = \tau(u_i) = 2M(M-1)\Omega$, and $\gamma(u_i) = 0$. We call the precinct nodes $v_i$ \textit{regular} and the precinct nodes $u_i$ \textit{auxiliary}. Let $k = \abs{\mathsf{V}}$, $c = 1/2$, and pick $\eps$ such that $0 < \eps < \frac{1}{M}$. Finally, let $\Pi = \{D_1, D_2, \ldots, D_k\}$, where $D_i = \{ v_i, u_i \}$ for all $i = 1, 2, \ldots, k$. Observe that we have $\beta(D_i) > \frac{1}{2}$ for all $i = 1, 2, \ldots, k$, i.e., we have $B_\Pi = V$ and $R_\Pi = \varnothing$. Note that since $0 < \eps < \frac{1}{M}$, we have $(2M^2 - 2M) \Omega < (1-\eps)2M^2 \Omega$ and $(2M^2 + 2M) \Omega > (1+\eps)2M^2 \Omega$, and thus every feasible district has total population exactly $2M^2 \Omega$.

Suppose the \ckspb\ instance is a $\texttt{Yes}$ instance, i.e., there is a subset $\mathsf{W} \subseteq \mathsf{V}$ of $M$ vertices such that its induced subgraph $\mathsf{H} \subseteq \mathsf{G}$ is connected, and $\sum_{\mathsf{v}_i \in \mathsf{W}} \omega(\mathsf{v}_i) \geq \Omega$. Let $W = \{v_i \mid \mathsf{v}_i \in \mathsf{W}\}$ be the set of regular precinct nodes corresponding to the vertices in $\mathsf{W}$. Then we have $\abs{W} = M$ and thus $\rho(W) = 2M^2 \Omega$. Furthermore, we have
\begin{align*}
    \sum_{v \in W \cap B_\Pi} \gamma(v) \tau(v) = \sum_{v \in W} \gamma(v) \rho(v) &= \sum_{i: \mathsf{v}_i \in \mathsf{W}} \Paren{\Paren{\frac{M-1}{2M}+\frac{\omega(\mathsf{v}_i)}{2\Omega}} \cdot 2M \Omega}\\
    &\geq \Paren{\frac{M-1}{2} + \frac{\Omega}{2\Omega}} \cdot 2M \Omega = M^2 \Omega = c \cdot \tau(W).
\end{align*}
Since $W$ induces a connected subgraph of $G$, it is a red $c$-deviating group of $\Pi$.

For the other direction, suppose the \audit\ instance is a $\texttt{Yes}$ instance, i.e., there is a red $c$-deviating group $W$ of $\Pi$. Suppose $W$ contains at least one auxiliary precinct node $u_i$. Recall that $\rho(W) = 2M^2 \Omega$. Hence, we have 
\[ \sum_{v \in W \cap B_\Pi} \gamma(v) \tau(v) = \sum_{v \in W} \gamma(v) \rho(v) < \sum_{v \in W \setminus \{u_i\}} \rho(v) = 2M \Omega \leq M^2 \Omega = c \cdot \tau(W), \]
a contradiction. Hence $W$ can only contain regular precinct nodes.
Then we have $\abs{\mathsf{W}} = M$, and
\begin{align*}
    \sum_{\mathsf{v}_i \in \mathsf{W}} \omega(\mathsf{v}_i) &= \sum_{v \in W \cap B_\Pi} \Paren{2\Omega \gamma(v) -\frac{(M-1)\Omega}{M}}\\
    &= \frac{1}{M} \cdot \sum_{v \in W} \Paren{\gamma(v) \rho(v) - (M-1)\Omega}\\
    &\geq \frac{\rho(W)}{M} - (M-1) \Omega = M \Omega - (M-1) \Omega = \Omega.
\end{align*}
Since $\mathsf{W}$ induces a connected subgraph of $\mathsf{G}$, the corresponding \ckspb\ instance is a $\texttt{Yes}$ instance.

Combining the above, there is a polynomial-time reduction from \ckspb\ to \audit. Since \audit\ is trivially in \textsf{NP}, we conclude that it is \textsf{NP}-complete.
\end{proof}

We then observe that the same reduction also gives the hardness of \generation.
\begin{theorem} \label{thm:gen_hardness}
\generation\ is \textsf{NP}-complete.
\end{theorem}
\begin{proof}
Observe that in each \audit\ instance constructed in the proof for Theorem~\ref{thm:audit_hardness}, the associated plan $\Pi$ is the only feasible redistricting plan. To see this, notice that every auxiliary node $u_i$ has a single neighbor $v_i$ and does not possess enough population to form a feasible district itself, and thus every $u_i$ must be paired with its corresponding $v_i$ to make a district. Therefore, the \audit\ and \generation\ are identical on this family of instances, and the hardness of \generation\ follows from the same reduction.
\end{proof}

\subparagraph{Remarks.} We note that the proofs above still hold even if we add more edges to the constructed graph. More specifically, for both the proof of Theorems~\ref{thm:audit_hardness} and~\ref{thm:gen_hardness}, we can safely add edges as long as (i) planarity is preserved and (ii) at least one of the endpoints of each additional edge is an auxiliary precinct node. To see this, observe that adding additional edges does not impact the feasibility of $\Pi$, and since valid deviating groups contain only regular precinct nodes, the set of possible deviating groups for $\Pi$ remains identical. For the proof of Theorem~\ref{thm:gen_hardness}, the additional edges may allow multiple feasible redistricting plans, but each of the feasible redistricting plans must still contain $k$ districts of one regular and one auxiliary precinct node each, and by the same reduction, either all or none of them are locally fair (corresponding to $\texttt{Yes}$ and $\texttt{No}$ \ckspb\ instances).

Note further that in both \audit\ and \generation, we do not explicitly require the districts and deviating groups to be compact with respect to any specific criteria. Under certain restrictive compactness constraints, the problems may become tractable. For example, if the districts and deviating groups are restricted to be subsets of precincts fully contained in a circle centered around a precinct point, then the set of possible districts and deviating groups has polynomial size, and thus \audit\ can be solved by enumeration in polynomial time. 

\section{Speeding up the DP and Sufficiency of Ensemble-based Auditing}
\label{s:dp_exp}
Although our dynamic programming algorithm for solving \audit\ on trees run in polynomial time, the time complexity of the algorithm is prohibitively high to be efficient in practice. Our goal in this section is to empirically demonstrate that this approach is not needed in practice, that is, it does not find reasonable deviating groups on plans that the ensemble-based method deems locally fair, hence showing that the ensemble-based auditing method is sufficient and obviating the need for the computationally inefficient dynamic programming.  

Towards this end, we first show that the dynamic program can be sped up significantly if (among other things) we {\em interpolate} the voter information to the entire population of a precinct, so that $\tau(v) = \rho(v)$ for each precinct $v$. After performing such interpolation on the data used in our experiments (Section~\ref{s:exp}), we first run the ensemble-based auditing method to find fair and unfair plans for $c = 0.5$. Next, for each of these plans, we run the dynamic program to find deviating groups, checking each one for compactness and the value $c$ for which it is a $c$-deviating group. We show that the dynamic program is unable to find compact deviating groups with $c \ge 0.52$ on the ensemble-audited $0.5$-locally fair plans (on the interpolated data). This demonstrates the the sufficiency of ensemble-based auditing if we relax the strength $c$ of the deviating group slightly.  

We note that our main experiments in Section~\ref{s:exp} use actual voter data, since it is unclear how such data should be interpolated in a principled way to the entire population. In the current section, we perform the interpolation in a simple way only to make the DP run efficiently, which in turn enables us to demonstrate the conceptual point that the ensemble-based method suffices. This provides strong evidence that even without interpolation, the ensemble-based method will suffice.

\subsection{Improving Running Time of Dynamic Program}
\label{ss:dp_details}
We first describe our approach to speed up the running time of the dynamic program.

\subparagraph{Special case of $\tau(v) = \rho(v)$.} We first assume that $\tau(v) = \rho(v)$ holds for all $v \in V$, i.e., every individual is labeled red or blue in every precinct. In this case, we can reduce the state space  by dropping the state variable $p$, that is, we let $A[v,i]$ denote the maximum number of unhappy blue voters in a subtree $W \subseteq T_v$ such that $v \in W$ and $\tau(W) = \rho(W) = i$. In this case, for leaf precincts $v$ we have
\[ A[v,i] = \begin{cases}
\text{uhp}(v), &\, i = \tau(v) = \rho(v);\\
0, &\, \text{otherwise,}
\end{cases}\]
and for general precincts $v$ (with children $\{u_1, \ldots, u_{\text{deg}(v)}\}$) we have
\begin{align}
    A[v,i] = \text{uhp}(v) + B_{v,i}[\text{deg}(v),p-\rho(v)], \label{vertical_dp_special_case}
\end{align}
where $B_{v,i}[1,x] = A[u_1,x]$ for all $x$, and for all $j \geq 2$ we have
\begin{align}
    B_{v,i}[j,x] = \max_{x' \in [0,x]} \left\{ B_{v,i}[j-1,x-x'] + A[u_j, x'] \right\}. \label{horizontal_dp_special_case}
\end{align}

Now, the algorithm computes $O(\abs{V} \cdot \sigma)$ values of $A[v,i]$, each computing $O(\mbox{deg}(T) \cdot \sigma)$ values of $B_{v,i}[j,x]$, each requiring $O(\sigma)$-time to loop through all values of $x'$. The overall time complexity thus drops to $O\Paren{\abs{V} \cdot \sigma^3 \cdot \mbox{deg}(T)}$. 

\subparagraph{Relaxing size of deviation.}
We next modify the state $A[v,i]$ to be the maximum number of unhappy blue voters in a subtree $W \subseteq T_v$ such that $v \in W$ and $\tau(W) = \rho(W) \leq i$, i.e., it is now allowed that the subtree $W$ has an aggregate population of less than $i$. Note that this induces a potential one-sided error in checking for the existence of deviating groups. To see this, consider the case when the algorithm finds some $(v,i)$ such that $i \in [(1-\eps)\sigma, (1+\eps)\sigma]$ and $A[v,i] > i/2$. Now this corresponds to a subtree $W$ rooted at $v$ with a population (or voter count) of \emph{at most} $i$ and a total number of unhappy blue voters of \emph{at least} $i/2$. While this still ensures a majority of voters in $W$ are unhappy voters of the same color, the actual population size may be less than $i$ and thus outside of the acceptable range $[(1-\eps)\sigma, (1+\eps)\sigma]$. However, we observe that this error is one-sided: Suppose there is indeed a deviating group $W$ of the correct population size $i \in [(1-\eps)\sigma, (1+\eps)\sigma]$ with a total number of unhappy blue voters of \emph{at least} $i/2$, our algorithm must either find $W$, or find a deviating group $W'$ with population at most $i$ and a larger number of unhappy blue voters. Therefore, if the algorithm does not find any deviating group under the relaxed definition, we can still conclude that there is no deviating group (with respect to the current spanning tree $T$).

\subparagraph{Pruning the states.} Under the modified semantics of $A[v,i]$, we observe that each $A[v,i]$ is non-decreasing in $i$ and each $B_{v,i}[j,x]$ is non-decreasing in $x$.  Now consider the computation of some fixed $B_{v,i}[j,x]$. We maintain an upper bound $ub$ and a lower bound $lb$ of $B_{v,i}[j,x]$. whenever $lb \geq ub$, we terminate the computation early and return $lb = ub$ as $B_{v,i}[j,x]$. Since the $B_{v,i}[j,x]$ are computed in increasing order of $x$, we initialize $lb = B_{v,i}[j,x-1]$ and let $lb = 0$ if $x = 0$. We also initialize $ub = B_{v,i}[j-1,x] + A[u_j, x]$. 

When the $\max$ function in Eq.~(\ref{horizontal_dp_special_case}) is evaluated in increasing order of $x'$, we update:
\begin{itemize}
    \item $lb \gets \max \{ lb, B_{v,i}[j-1,x-x'] + A[u_j, x'] \}$;
    \item $ub \gets \min \{ ub, B_{v,i}[j-1,x-x'] + A[u_j, x] \}.$ 
    \end{itemize}
The second step  is because that for any $x'' > x'$, we have
    $$B_{v,i}[j-1,x-x''] + A[u_j, x''] \leq B_{v,i}[j-1,x-x'] + A[u_j, x].$$
Therefore, $B_{v,i}[j-1,x-x'] + A[u_j, x]$ is the maximum possible value of $B_{v,i}[j,x]$ if the function is maximized at any $x''>x'$. If this is matched by $lb$, then the final maximum value will be exactly $lb$. In this case, we terminate the computation without examining any $x''>x'$ in Eq.~(\ref{horizontal_dp_special_case}).

The same idea is applied to the computation of $A[v,i]$: We maintain a lower bound $lb'$ for $A[v,i]$ (initialized to $A[v, i-1]$), and whenever 
$$lb' \geq ub' = \text{uhp}(v) + \sum_{j=1}^{\text{deg}(v)} A[u_j, i'],$$ 
we return $A[v,i] = lb'$. 

\paragraph{Rounding the population.} 
For a fixed threshold parameter $P$, we round each $\rho(v)$ down to the largest multiple of $P$ that is smaller than or equal to $\rho(v)$. Formally, let $\rho'(v) = \floor{\frac{\rho(v)}{P}} \cdot P$. For any subtree $W \subseteq T$, we have $\sum\limits_{v \in W} \rho'(v) \leq \sum\limits_{v \in W} \rho(v).$

Let $A'[v,i]$ denote the output of the algorithm when the rounded population level $\rho'(v)$ is used instead of $\rho(v)$. Then we have $A'[v,i] \geq A[v,i]$. Therefore, running our algorithm with rounding introduces one-sided error: If there exists any deviating group $W$ of $\Pi$ of population level $i$, then the algorithm with rounding can also output $W$ since it has the same number of unhappy blue voters with a rounded-down population level. We must then relax the acceptable size range from $[(1-\eps)\sigma, (1+\eps)\sigma]$ to $[(1-\eps)\sigma - P\cdot|V|, (1+\eps)\sigma]$ to accommodate this error and so that $W$ becomes a candidate deviating group. 

\paragraph{Final running time.} With all these strategies incorporated, the running time of the dynamic program is reduced to $O\Paren{\abs{V} \cdot (\frac{\sigma}{P})^3 \cdot \mbox{deg}(T)}$ as there are now only $O\Paren{\frac{\sigma}{P}}$ population levels. This is the version of the algorithm that we implement.

\subsection{Empirical Results and Sufficiency of Ensemble-based Auditing}
\label{ss:exp_dp}
We now use both our ensemble approach and the dynamic program to audit the ensemble for NC. We use the same experimental setup as in Section~\ref{s:exp}, except for one change. Since the DP assumes $\tau(v) = \rho(v)$, we need to interpolate the voter labels to the entire precinct. We do this in the natural way. We keep the same $\gamma(v)$ and $\beta(v)$ values determined from an election, but let $\tau(v) = \rho(v)$. The number of red and blue voters in a precinct $v$ become $\gamma(v) \cdot \rho(v)$ and $\beta(v) \cdot \rho(v)$, respectively. This is equivalent to assuming that in each precinct $v$, the rate of red/blue preferences of non-voters is identical to that of the voters. Accordingly, a $c$-deviating group must have a $c$ fraction of the total population being unhappy individuals of the same color. 

Using the interpolated voter labels on NC data, we first run the ensemble-based auditing method assuming $c = 0.5$. We find that 52 among the 1,000 plans ($5.2\%$) in the ensemble do not have $0.5$-deviating groups and are deemed fair by the ensemble approach. We again rank the plans in the ensemble by their unfairness score $\unf(\Pi)$. We then construct two groups of plans: (1) 26 Plans deemed $0.5$-fair by the ensemble approach; (2) 10 Plans in the bottom 5\% (most unfair, in terms of $\unf$ score) in the ranking. We generate $5$ random spanning trees of the NC precinct graph. For each plan and each spanning tree, we run the dynamic program where the population rounding parameter is set as $P = 750$. For each group of plans (fair and unfair), we obtain the set of all deviating groups found by the dynamic program on any spanning tree. We measure the Polsby-Popper compactness score of each deviating group on the original graph and the {\em strength} $c$ for which the group is a $c$-deviating group in that plan (the largest $c$ for which the group is indeed deviating). Note that a larger value of $c$ implies the deviation is robust to small population changes, and is more significant in terms of unfairness. 

\begin{figure}[htbp]
   \begin{subfigure}[b]{0.4\textwidth}
    \centering
    \includegraphics[width=\textwidth]{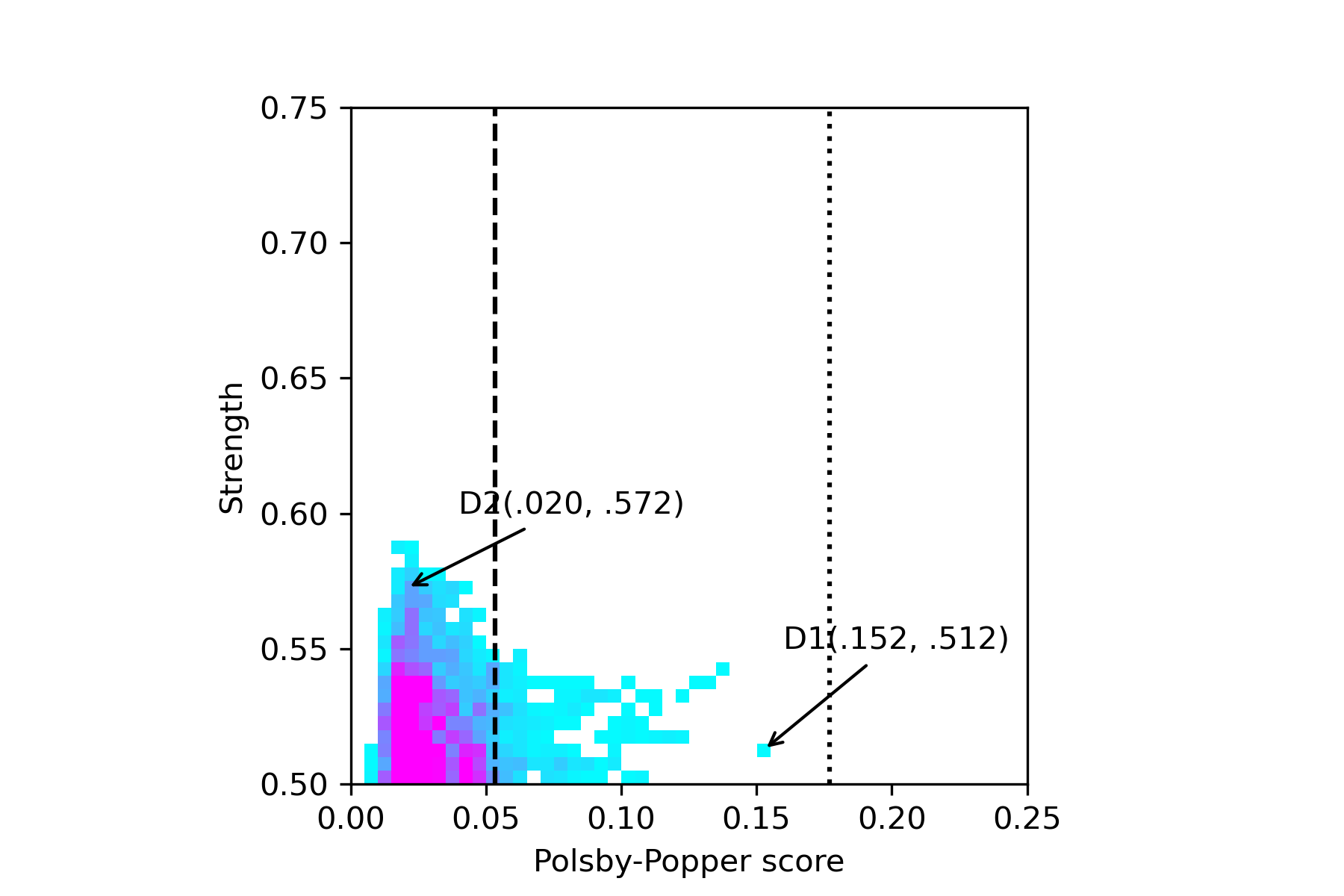}
    \caption{Fair plans.}
    \label{fig:heatmap_good_plans}
\end{subfigure}
\hfill
   \begin{subfigure}[b]{0.4\textwidth}
    \centering
    \includegraphics[width=\textwidth]{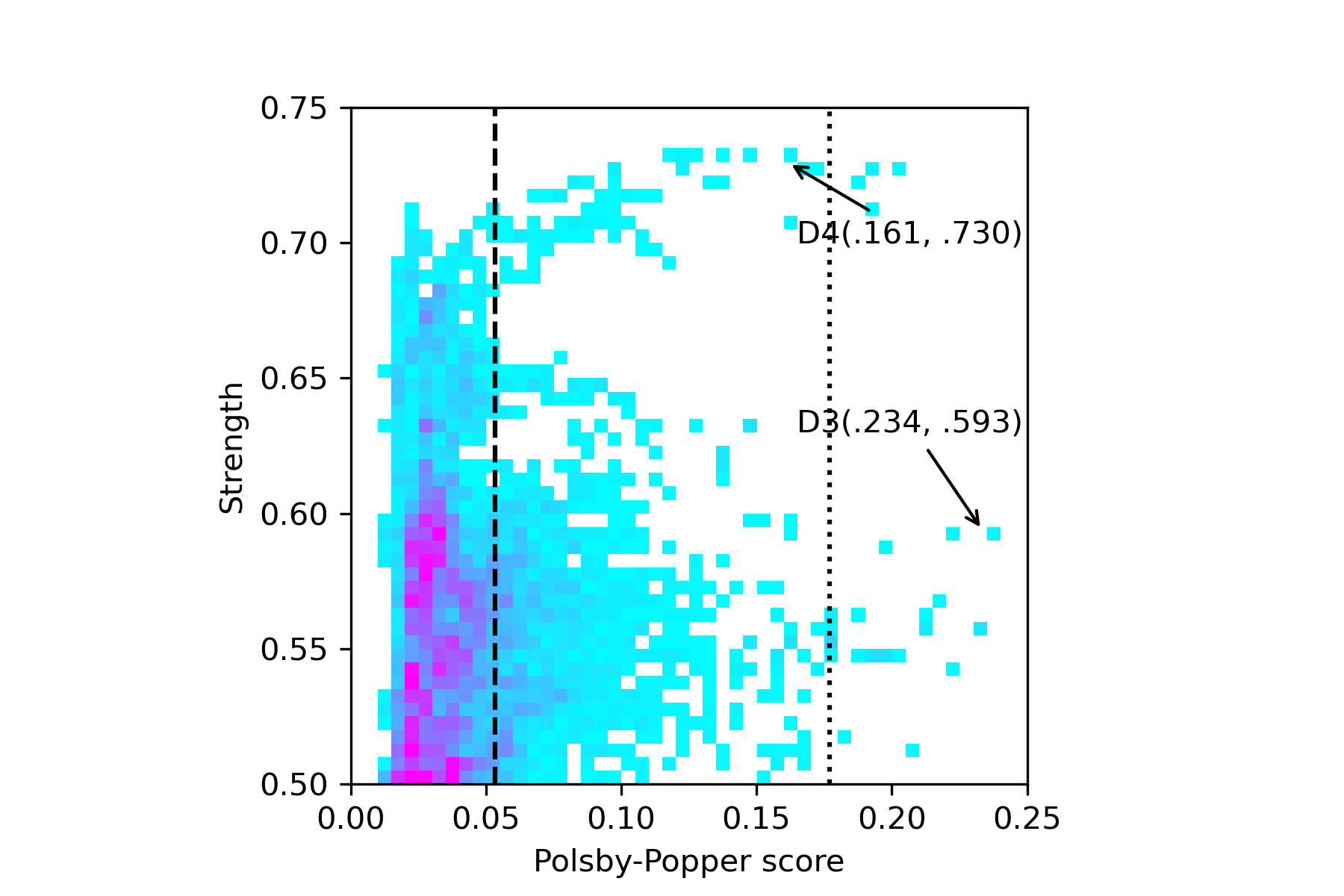}
    \caption{Unfair plans.}
    \label{fig:heatmap_bad_plans}
\end{subfigure}
\caption{Heatmaps of deviating groups found by the dynamic program on fair and unfair NC plans.}
\end{figure}

\begin{figure}[htbp]
   \begin{subfigure}[b]{0.24\textwidth}
    \centering
    \includegraphics[width=\textwidth]{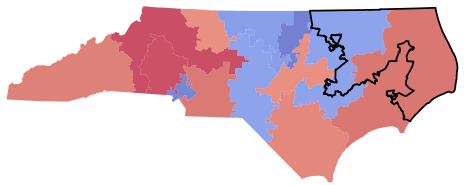}
    \caption{A $.512$-deviating group with a $.152$ Polsby-Popper score ($D1$ in Fig.~\ref{fig:heatmap_good_plans}).}
    \label{fig:weak_dg}
\end{subfigure}
\hfill
   \begin{subfigure}[b]{0.24\textwidth}
    \centering
    \includegraphics[width=\textwidth]{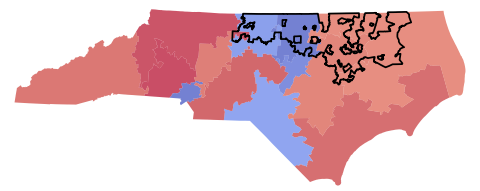}
    \caption{A $.572$-deviating group with a $.020$ Polsby-Popper score ($D2$ in Fig.~\ref{fig:heatmap_good_plans}).}
    \label{fig:bogus_dg}
\end{subfigure}
\hfill
   \begin{subfigure}[b]{0.24\textwidth}
    \centering
    \includegraphics[width=\textwidth]{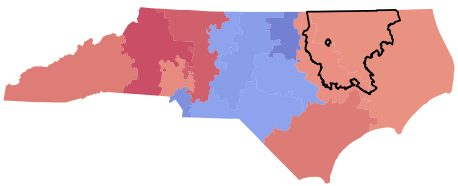}
    \caption{A $.593$-deviating group with a $.234$ Polsby-Popper score ($D3$ in Fig.~\ref{fig:heatmap_bad_plans}).}
    \label{fig:strong_dg}
\end{subfigure}
\hfill
   \begin{subfigure}[b]{0.24\textwidth}
    \centering
    \includegraphics[width=\textwidth]{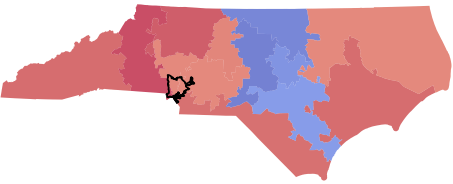}
    \caption{A $.730$-deviating group with a $.161$ Polsby-Popper score ($D4$ in Fig.~\ref{fig:heatmap_bad_plans}).}
    \label{fig:compact_dg}
\end{subfigure}
\caption{Deviating groups found by the dynamic program for two fair plans (left two maps) and two unfair plans (right two maps).}
\end{figure}

In Figures~\ref{fig:heatmap_good_plans} and~\ref{fig:heatmap_bad_plans}, we plot the heatmaps of the deviating groups found by the dynamic program for the fair and unfair sets of plans, respectively, where the $x$-axis and $y$-axis demonstrate their Polsby-Popper score and their \emph{strength} respectively. As shown, for the plans deemed $0.5$-fair by the ensemble approach, most deviating groups are either not compact (having low Polsby-Popper scores of $<0.1$) or not strong (having strength values close to $0.5$). As context, the minimum and average Polsby-Popper scores over all NC districts in the \recom-generated NC ensemble are $0.053$ and $0.177$, respectively (corresponding to the two vertical lines in both plots); in other words, deviating groups with Polsby-Popper scores of $<0.053$ (to the left of the dashed vertical line) are less compact than \emph{every one} of the 10k districts in the ensemble. 

In fact, our dynamic program finds no deviating group with an above-average Polsby-Popper score for any $0.5$-fair plan. The closest deviating group, shown as D1 in Figure~\ref{fig:heatmap_good_plans}, has Polsby-Popper score $0.152$ and a strength of $0.512$; we visualize it in Figure~\ref{fig:weak_dg}. In Figure~\ref{fig:bogus_dg}, we visualize another deviating group found by the dynamic program (shown as D2 in Figure~\ref{fig:heatmap_good_plans}) with a Polsby-Popper score of $0.020$ and a strength of $0.572$. As manifested in the visualizations, deviating groups with very low Polsby-Popper scores like $0.02$ are spurious with artificial shapes (such as holes) and should not be considered when it comes to determining whether a redistricting plan is fair. All the deviating groups for the $0.5$-fair plans with a Polsby-Popper score at least $0.053$ have lower strength ($<0.55$). In summary, we can reasonably conclude that most of the fair plans found by the ensemble-based auditing approach do not admit strong, contiguous, and reasonably compact deviating groups even when audited by the dynamic program.

In contrast, for the plans ranked in the bottom $5\%$ according to the ensemble-based approach, the dynamic program is able to find both strong and compact deviating groups quite easily. In Figures~\ref{fig:strong_dg} and~\ref{fig:compact_dg} we show (a) a $.593$-deviating group with a $.234$ Polsby-Popper score (D3 in Figure~\ref{fig:heatmap_bad_plans}), and (b) a $.730$-deviating group with a $.161$ Polsby-Popper score (D4 in Figure~\ref{fig:heatmap_bad_plans}) that the dynamic program find for one of the unfairest plans. These results show that the strengths of deviating groups for fair plans (according to ensemble based auditing) are considerably lower than that for unfair plans. In other words, the results via DP validates that via the ensemble based approach.
  

\section{Alternative Fairness and Compactness Metrics}
\label{ss:alternative_exps}

\begin{figure}[htbp!]
    \begin{subfigure}[b]{0.98\textwidth}
        \centering
        \includegraphics[width=\textwidth]{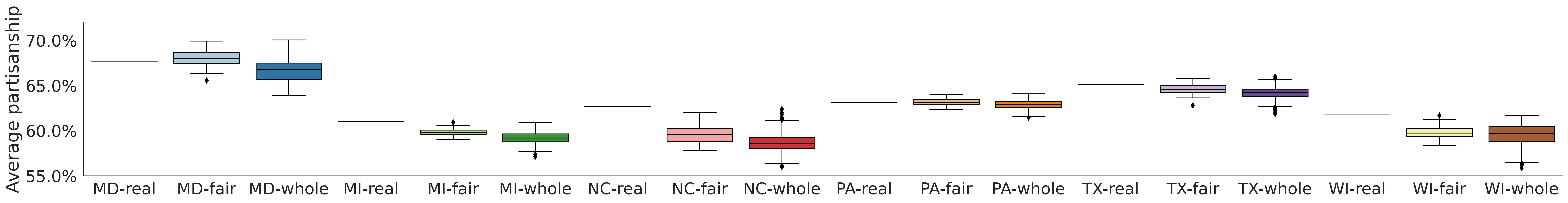}
        \caption{Average partisanship}
        \label{fig:partisanship}
    \end{subfigure}
    \begin{subfigure}[b]{0.98\textwidth}
        \centering
        \includegraphics[width=\textwidth]{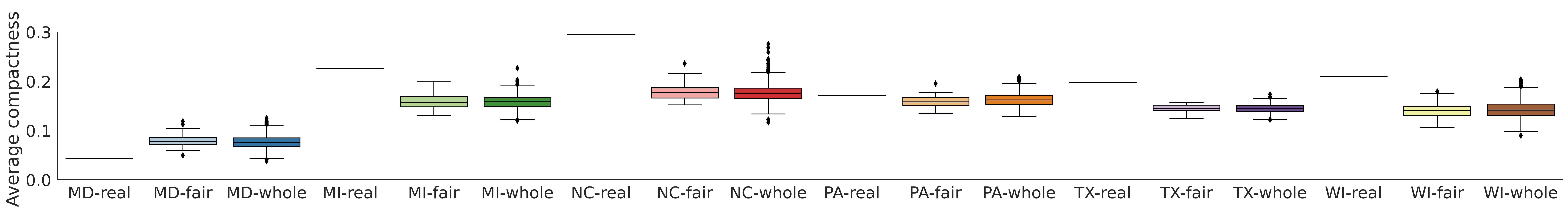}
        \caption{Average Polsby-Popper score}
        \label{fig:compactness_avg}
    \end{subfigure}
    \caption{Distribution of alternative fairness and compactness metrics among subsets of generated plans.}
\end{figure}

\noindent {\bf Average partisanship.} 
For each district, let its \emph{partisanship} be the percentage of votes in the majority color.
Therefore, a low partisanship (towards 50\%) implies better competitiveness in that district. 
We define the average partisanship of $\Pi$ to be the average of partisanship values over its districts (ignoring small differences in population) as an alternative to the competitiveness metric used in Figure~\ref{fig:competitiveness}. In Figure~\ref{fig:partisanship}, we compare the average partisanship among the three subsets of plans used in Section~\ref{ss:compatibility} (top-5\% fairest plans, whole ensemble, and real enacted plans).

Results show that fair plans generate slightly more partisan districts. However, compared to the whole ensemble, the (roughly) 2-3\% of shift in average partisanship is small and comparable to other uncertainties (e.g., voter turnouts or year-to-year election result gaps). Furthermore, the median average partisanship of the fair plans is smaller than that of the real-world redistricting plan for all but one state, showing that local fairness remains compatible with reasonably small partisanship.

\noindent {\bf Average compactness.} 
We define the average compactness of $\Pi$ to be the average of Polsby-Popper scores over its districts. In Figure~\ref{fig:compactness_avg} we compare the average compactness among the three subsets of plans. Similar to the results for minimum compactness, the average compactness of the locally fair plans remains comparable to that of the entire ensemble. On the other hand, the enacted plans perform better on average compactness than on minimum compactness, showing that enacted plans have larger variances in the compactness scores than plans in the ensemble.

\section{Robustness of Local Fairness to Voting Patterns}
\label{s:additional_exp}
\label{ss:robustness}

To test the robustness of the local fairness notion to changes in voting patterns, we repeat the ensemble-based audit process in Section~\ref{s:exp} for MD, NC, PA, and TX with $\gamma(v), \tau(v)$, and $\beta(v)$ values replaced by label values obtained from the $2012$ presidential election. We do not consider MI (only a few plans are fair, so the sample size is too small) and WI (most plans are fair, and thus the ensemble and fair plans yield similar statistics). 

\begin{figure}[t]
\centering
    \begin{subfigure}[t]{0.98\textwidth}
    \centering
    \includegraphics[width=\textwidth]{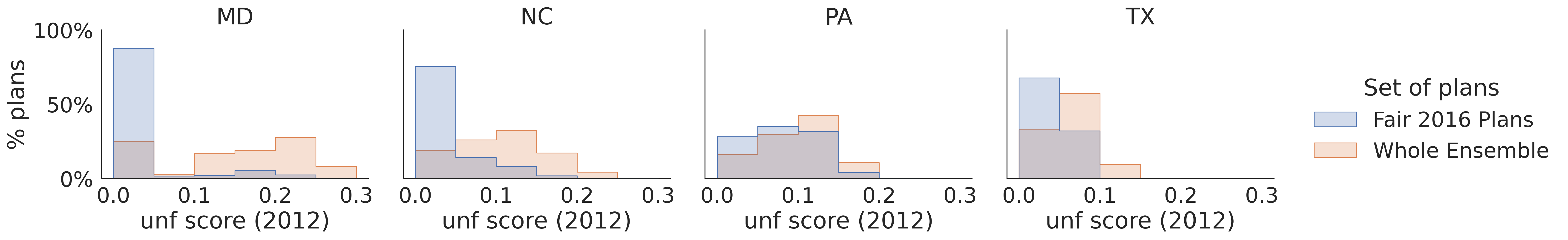}
    \end{subfigure}
    \\
    \begin{subfigure}[t]{0.98\textwidth}
    \centering
    \includegraphics[width=\textwidth]{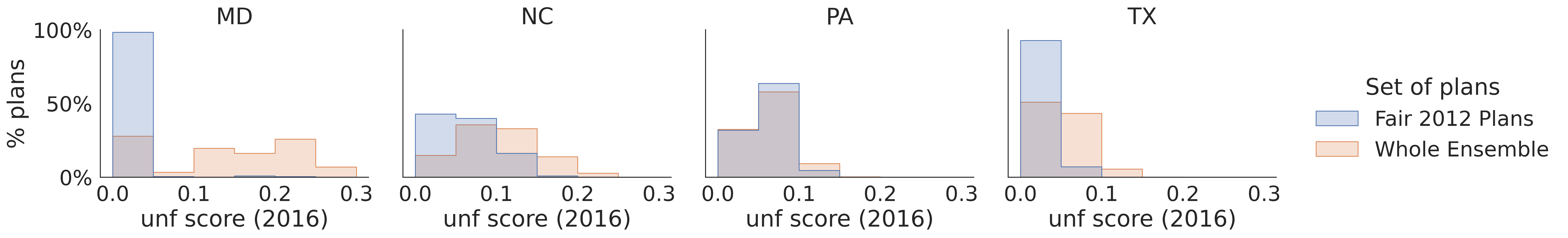}
    \end{subfigure}    

\caption{Histograms of $\unf$ scores. The top plots compute $0.5$-fair maps using 2016 voter data, and the blue bars plot the histograms of $\unf$ scores when these plans are audited using 2012 voter data. The orange bars represent the histograms of $\unf$ scores of the entire ensemble when audited using 2012 data. The bottom plots switch the roles of 2016 and 2012, so the $0.5$-fair plans are generated using 2012 data and audited using 2016 data.}
\label{fig:12-16}
\end{figure}

For $c=0.5$, we obtain the set of locally fair plans (i.e., plans without $c$-deviating groups) when audited using 2012 (resp. 2016) voter labels. We then compute the unfairness of these plans using the 2016 (resp. 2012) voter data, i.e., from the other election. We repeat this for all the plans in the ensemble, obtaining the $\unf$ score for each plan.  We plot the histograms of these $\unf$ values in Figure~\ref{fig:12-16}, where the x-axis is the bucketed $\unf$ score, and the $y$-axis is the percentage of plans among the fair maps (resp. ensemble) that fall in that bucket. 

For MD, NC, and TX, the blue bars are skewed significantly towards the left compared to the entire ensemble, showing that the fairest plans identified by auditing with a specific election remains significantly fairer compared to the entire ensemble when measured by another election. This shows the local fairness notion is fairly robust, or insensitive to year-to-year election result fluctuations. For PA, the fair plans are more sensitive to the specific election used, which reflects the role of PA as a swing state across elections. 

\section{Visualization of Fair and Unfair Plans for Other States}
\label{ss:visualization}
\begin{figure}[!htbp]
    \begin{subfigure}[t]{0.3\textwidth}
    \centering
    \includegraphics[width=\textwidth]{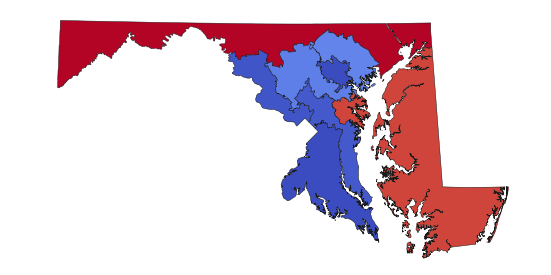}
    \caption{MD fair plan}
    \label{fig:md_fair_map}
    \end{subfigure}
      \hfill
       \begin{subfigure}[t]{0.3\textwidth}
    \centering
    \includegraphics[width=\textwidth]{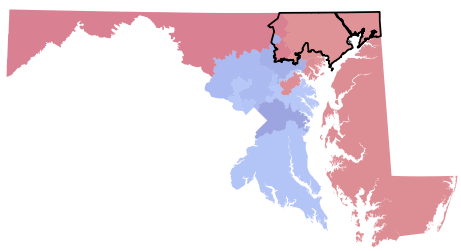}
    \caption{MD plan with a blue deviating group}
    \label{fig:md_dg_ex}
    \end{subfigure}
      \hfill
    \begin{subfigure}[t]{0.3\textwidth}
    \centering
    \includegraphics[width=\textwidth]{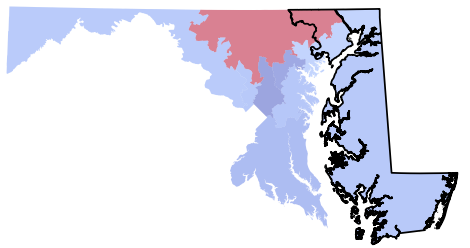}
    \caption{MD plan with a red deviating group}
    \label{fig:md_dg_ex2}
    \end{subfigure}
    \caption{Maryland plans without and with deviating groups}
\end{figure}

\begin{figure}[!htbp]
    \begin{subfigure}[t]{0.25\textwidth}
    \centering
    \includegraphics[width=\textwidth]{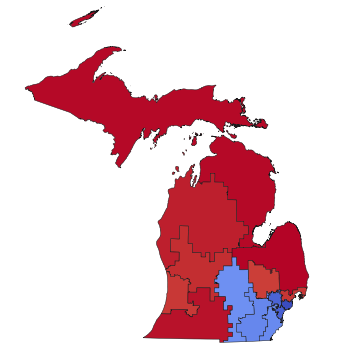}
    \caption{MI fair plan}
    \label{fig:mi_fair_map}
    \end{subfigure}
      \hfill
       \begin{subfigure}[t]{0.25\textwidth}
    \centering
    \includegraphics[width=\textwidth]{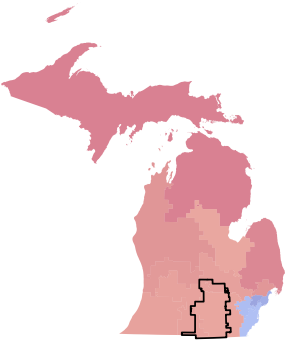}
    \caption{MI plan with a blue deviating group}
    \label{fig:mi_dg_ex}
    \end{subfigure}
      \hfill
    \begin{subfigure}[t]{0.25\textwidth}
    \centering
    \includegraphics[width=\textwidth]{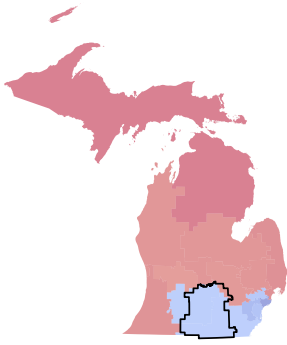}
    \caption{MI plan with a red deviating group}
    \label{fig:mi_dg_ex2}
    \end{subfigure}
    \caption{Michigan plans without and with deviating groups}
\end{figure}

\begin{figure}[t]
    \begin{subfigure}[t]{0.2\textwidth}
        \centering
        \includegraphics[width=\textwidth]{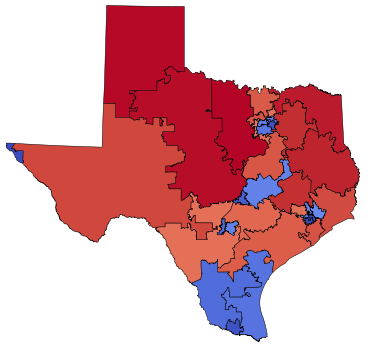}
        \caption{TX fair plan}
        \label{fig:tx_fair_map}
    \end{subfigure}
       \hfill
    \begin{subfigure}[t]{0.2\textwidth}
        \centering
        \includegraphics[width=\textwidth]{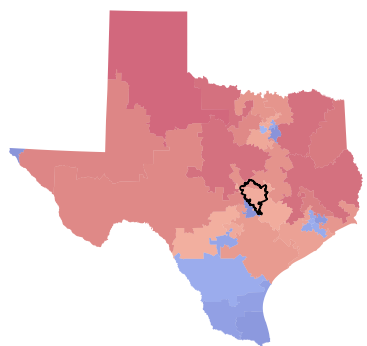}
        \caption{TX plan with a blue deviating group}
        \label{fig:tx_dg_ex}
    \end{subfigure}
        \hfill
    \begin{subfigure}[t]{0.2\textwidth}
        \centering
        \includegraphics[width=\textwidth]{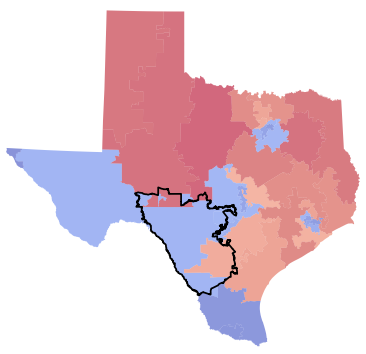}
        \caption{TX plan with a red deviating group}
        \label{fig:tx_dg_ex2}
    \end{subfigure}
    \caption{Texas plans without and with deviating groups}
\end{figure}

\begin{figure}[h]
    \begin{subfigure}[t]{0.2\textwidth}
    \centering
    \includegraphics[width=\textwidth]{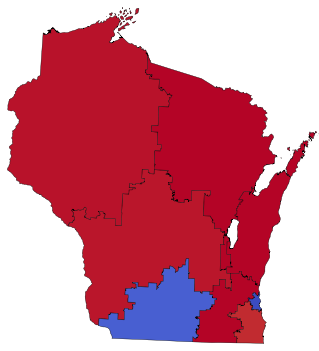}
    \caption{WI fair plan}
    \label{fig:wi_fair_map}
    \end{subfigure}
      \hfill
       \begin{subfigure}[t]{0.2\textwidth}
    \centering
    \includegraphics[width=\textwidth]{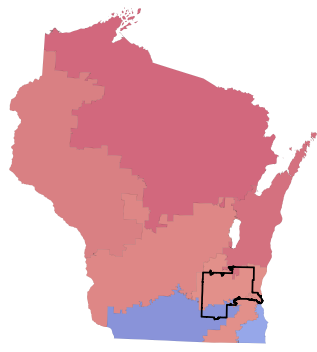}
    \caption{WI plan with a blue deviating group}
    \label{fig:wi_dg_ex}
    \end{subfigure}
      \hfill
    \begin{subfigure}[t]{0.2\textwidth}
    \centering
    \includegraphics[width=\textwidth]{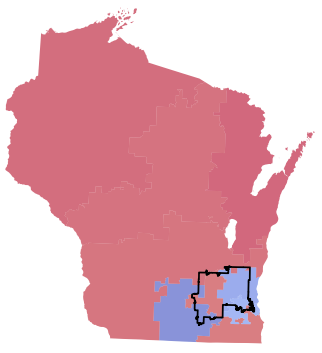}
    \caption{WI plan with a red deviating group}
    \label{fig:wi_dg_ex2}
    \end{subfigure}
    \caption{Wisconsin plans without and with deviating groups}
\end{figure}

We show additional visualizations of fair plans and deviating groups found using ensemble-based auditing. 
As before, we show deviating groups with black outline, and the districts are coded by its color and the extent of partisanship: districts with a larger value of $\gamma$ (resp. $\beta$) are colored in darker red (resp. blue).
For each state (MD, MI, TX, WI), we show a fair plan (Figures~\ref{fig:md_fair_map},~\ref{fig:mi_fair_map},~\ref{fig:tx_fair_map},~\ref{fig:wi_fair_map}), and an example of a deviating group of each color. We discuss the deviating groups in each state. 

In MD, the central blue districts tend to not be competitive, and the geography of the state contributes to the difficulty of forming red deviating groups. Thus MD is the only state where the precincts in the most deviating groups are not densely populated areas (see Figure~\ref{fig:md_precincts}). Instead, the two ``panhandles'' (the western and coastal eastern regions of the state) tend to be part of deviating groups, see Figures~\ref{fig:md_dg_ex} and~\ref{fig:md_dg_ex2}. 

In Figures~\ref{fig:mi_dg_ex} and~\ref{fig:mi_dg_ex2} we show a red and a blue deviating group in MI. Michigan had the lowest number of fair plans in the ensemble (see Table~\ref{tab:achievability}). The precincts in deviating groups are clustered in one region of MI, where the districting is sensitive to which precincts belong in red or blue districts. Both red and blue deviating groups are concentrated around this area.

In contrast, some deviating groups in TX concentrate around urban areas, while others intersect a large area of less densely populated precincts. The districting shown in Figure~\ref{fig:tx_dg_ex} shows a blue deviating group in proximity to an urban area. In contrast, Figure~\ref{fig:tx_dg_ex2} shows a red deviating group on a districting plan intersecting a large blue district. 

In Figure~\ref{fig:wi_dg_ex}, a blue deviating group of WI pulls in a portion of an urban area (left) and spans across to another blue district. In the fair plan, the urban counties are contained fully in a blue district, while the middle red precincts that deviate in Figure~\ref{fig:wi_dg_ex2} are in their own red district. 

\end{document}